\pdfoutput=1
\documentclass[11pt]{article}

.5 scaled \magstep4
\font\medio=cmr9.5 scaled \magstep2
\outer\def\beginsection#1\par{\medbreak\bigskip
      \message{#1}\leftline{\bf#1}\nobreak\medskip
\vskip-\parskip
      \noindent}
\usepackage{graphicx} 
\begin{document}
\bibliographystyle {unsrt}

\begin{center}
{\Large {\bf Flat spectra of cosmic gravitons in the nHz and audio bands}}\\
\vspace{15mm}
 Massimo Giovannini 
 \footnote{Electronic address: massimo.giovannini@cern.ch} \\
\vspace{1cm}
{{\sl Department of Physics, CERN, 1211 Geneva 23, Switzerland }}\\
\vspace{0.5cm}
{{\sl INFN, Section of Milan-Bicocca, 20126 Milan, Italy}}

\vspace*{1cm}
\end{center}

\centerline{\medio  Abstract}
\vspace{5mm}
The spectra of the relic gravitons are customarily normalized in the low-frequency domain where the signal of the concordance paradigm is expected to peak and this is why their contribution to the temperature and polarization anisotropies of the microwave background is only described by the tensor to scalar ratio. If the consistency relations are broken, the same strategy is accomplished by introducing the tensor spectral index as a further independent parameter. When the dominant component of the spectral energy density is distributed for frequencies much larger than the aHz, the logic behind this conventional approach is much less compelling. The improved bounds in the audio band and the current data from the pulsar timing arrays 
in the nHz region motivate a new strategy for the absolute normalization of the cosmic background of relic gravitons. After introducing a general four-dimensional action for the analysis of the relic gravitons the new approach is illustrated in the case of conventional and unconventional inflationary models. 
\vskip 0.5cm

\nonumber
\noindent

\vspace{5mm}

\vfill
\newpage

\renewcommand{\theequation}{1.\arabic{equation}}
\setcounter{equation}{0}
\section{Introduction}
\label{sec1}
Since the evolution of the tensor modes of the geometry is not Weyl-invariant \cite{gr1,gr2,par1} the production of relic gravitons is expected, with different phenomenological signatures, in a variety of scenarios and, in particular, during an isotropic phase of quasi-de Sitter expansion \cite{star1}. Besides the vanilla $\Lambda$CDM paradigm\footnote{$\Lambda$ stands for the dark-energy component while the CDM denotes the Cold Dark Matter contribution. The simplest scenario where the neutrinos are massless, the dark-energy does not fluctuate and the tensor modes are absent is customarily referred to as the vanilla $\Lambda$CDM model. }, the simplest version of the concordance scenario includes only one further free parameter, namely the ratio $r_{T}(k_{p})$  describing the tensor component of the large-scale inhomogeneity at a conventional pivot scale (see e.g. \cite{RT1,RT2,RT3}) that coincides, in what follows, with $k_{p} = 0.002\,\, \mathrm{Mpc}^{-1}$. This scale corresponds\footnote{The units $\hbar= c =1$ will be consistently employed throughout. Furthermore in this investigation the scale factor is normalized as $a(\tau_{0}) = a_{0} =1$. Finally, the standard prefixes shall be consistently employed, e.g.   $1\,\mathrm{aHz} = 10^{-18}, \mathrm{Hz}$, 
$1\, \mathrm{mHz} = 10^{-3}\, \mathrm{Hz}$,  $1\,\mathrm{MHz} = 10^{6} \, \mathrm{Hz}$ and so on and so forth. } to a comoving frequency $\nu_{p} = k_{p}/(2\pi) = 3.09\,\,\mathrm{aHz}$ and this is why, in the conventional lore, the limits on $r_{T}(k_{p})$ translate into constraints on the spectral energy density of the relic gravitons in the aHz range. The spectral energy density in critical units (and at the present time) is  denoted hereunder by $\Omega_{gw}(\nu) \equiv \Omega_{gw}(\nu,\tau_{0})$ where $\nu$ is the comoving frequency. In the concordance paradigm $\Omega_{gw}(\nu)$
approximately scales as $\nu^{-2}$ between the $\mathrm{aHz}$ and $100 \,\mathrm{aHz}$ \cite{ir1,ir2,ir3} while it is flat (or slightly decreasing) for larger frequencies. 

Two tacit assumptions are implicitly associated with $r_{T}$: the first one is that the early completion of the concordance paradigm is provided by the conventional inflationary scenario, the second is that the consistency relations are not violated\footnote{For the sake of conciseness the argument of $r_{T}(k)$ will be dropped when not strictly necessary and we shall therefore employ the following shorthand notation  $r_{T}= r_{T}(k_{p})$.}. Between these two assumptions (which are rarely stressed) the former seems stronger than the latter but they are instead equally essential if the only final objective is a stringent set of bounds on $r_{T}$ from the temperature and polarization anisotropies of the microwave background. In practice the consistency relations stipulate that the tensor spectral index $n_{T}$ and the slow-roll parameter $\epsilon$ are both determined by the value of $r_{T}$ according to the following (approximate) chain  of equalities $ n_{T} \simeq - r_{T}/8 \simeq - 2 \epsilon$. The consistency relations are valid in the case of single-field inflationary 
scenarios (see e.g. \cite{wein}) but they can be otherwise broken. Since the vanilla $\Lambda$CDM only represents a useful compromise between the available data and the number of ascertainable parameters, the addition of a tensor component (only described by $r_{T}$) allows for a rather accurate set of limits implying, in a conservative perspective, that $r_{T} \leq 0.06$ \cite{RT1,RT2,RT3}. If the consistency conditions are broken the accuracy on $r_{T}$ becomes comparatively smaller and, for this reason, some time ago it has been suggested that the bounds on $r_{T}$ must be viewed in a broader perspective where the analysis of the three standard cosmological data sets (i.e. cosmic microwave background anisotropies, large-scale structure and supernovae), is combined with the bounds of wide-band interferometers in the so-called audio band (i.e. between few Hz and few kHz) \cite{norm}. 

Since direct measurements are now available it seems appropriate to reconsider the high-frequency normalization of the cosmic backgrounds of relic gravitons 
as a possible alternative to the conventional approach based on the analysis of 
the aHz region. There are two complementary motivations corroborating this suggestion and they are associated with the recent claims of the pulsar timing arrays (PTA) in the nHz band and with the improved constraints provided by the 
wide-band interferometers in the audio band. Indeed, in the last thirty years the pulsars 
provided a series of relevant upper limits on the spectral energy density 
of the relic gravitons at intermediate frequencies\cite{PUL1,PUL2,PUL3,PUL4,PUL5}. Broadly speaking 
the previous results\footnote{As usual $h_{0}$ is the Hubble rate expressed in units of $100\,\mathrm{Hz}\, \mathrm{km}/\mathrm{Mpc}$ and since $\Omega_{gw}(\nu)$
denotes the spectral energy density in critical units, $h_{0}^2$ appears 
in its denominator. For this reason it is common practice to phrase the discussions directly in terms of  $h_{0}^2 \Omega_{gw}(\nu)$ that is independent of the specific value of $h_{0}$.} suggested $h_{0}^2\,\Omega_{gw}(\nu,\tau_{0}) \, <\, 10^{-10}$ 
for $\nu = \nu_{P} = {\mathcal O}(\mathrm{nHz})$. More recently the PTA
reported a series of effects that could be attributed to the relic gravitons \cite{NANO1,PPTA,EPTA,IPTA}. Barring for the slight differences between results 
reported by the four collaborations, the current evidence might suggest that $h_{0}^2 \Omega_{gw}(\nu,\tau_{0}) = {\mathcal O}(10^{-9})$ for a narrow slice of frequencies approximately ranging between a fraction of the nHz and $30$ nHz. The observations of Refs. \cite{NANO1,PPTA,EPTA,IPTA}  are, at the moment, very preliminary and the key property of a PTA is that the signal from relic gravitons, unlike a potential noise, is correlated across the baselines.  These correlations have not been observed but it is nonetheless interesting to consider more carefully the relic gravitons in the nHz domain; this is the perspective already conveyed some years ago \cite{REFR1,REFR2} before the 
evidences of the PTA.  The potential signals in the nHz band must anyway be complemented by a series of more robust limits in the audio band coming from the direct measurements of wide-band detectors. Depending  on the slope of the spectral energy density,  
the bounds from wide-band interferometers slowly improved through the years \cite{STone,STonea,STtwo,STthree,STthreea,STfour,STfive} and finally led to the joint analysis of the LIGO, Virgo and KAGRA collaborations \cite{STsix} suggesting that we can even have $\Omega_{gw}(\nu,\tau_{0})  \leq {\mathcal O}(10^{-9})$ for typical frequencies between $20$ Hz and $80$ Hz. The current measurements in the nHz and in the audio band seem then to point towards a quasi-flat spectral energy density of the relic gravitons when the comoving 
frequency encompasses the nHz and the audio bands.

In this paper we argue that even in the context of conventional inflationary models, the spectral energy density of the relic gravitons can be quasi-flat at high-frequency with typical amplitudes much larger than the ones of the concordance scenario. The potential signal must then be compatible, in this context, with the PTA observations and consistent with the limits of wide-band interferometers in the audio band. Instead 
of worrying about $r_{T}$ (as customarily done in the standard approach \cite{RT1,RT2,RT3}) we can directly impose the high-frequency 
normalization\footnote{The approach discussed here does 
not exclude that $r_{T}$ is drastically smaller than the current bounds stemming from the temperature and 
the polarization anisotropies of the microwave background. }. To pursue this possibility 
we first propose a general four-dimensional action for the consistent analysis of the relic gravitons evolving in conformally flat background geometries and generalizing the Ford-Parker action of Ref. \cite{par1}.
After introducing the different parametrizations of the action and its quantization, 
the conditions leading to a flat slope at high-frequency are analyzed by computing 
the spectral energy density within the Wentzel-Kramers-Brillouin (WKB).
The strategy for the high-frequency normalization is then
explained by considering the PTA evidences, the limits from the wide-band 
detectors and the other constraints customarily associated with the high-frequency gravitons. In short the layout of the paper is the following. In section \ref{sec2} the basic action of the problem is analyzed in its different forms. As suggested by the current data, the normalization of the spectral energy density in the nHz and audio bands is discussed in section \ref{sec3}. Section \ref{sec4} is focused on the conditions for a flat  high-frequency spectrum with amplitudes potentially much larger than the signal of the vanilla $\Lambda$CDM scenario. In section \ref{sec5} we present some phenomenological 
considerations that complement the results of section \ref{sec4}. Finally section \ref{sec6} 
contains the concluding remarks. 

\renewcommand{\theequation}{2.\arabic{equation}}
\setcounter{equation}{0}
\section{General parametrization of the action}
\label{sec2}
\subsection{The general action and its parametrizations}
In the single-field case the effective action of generic inflationary models involves all the terms that include four derivatives and are suppressed by the negative powers of a large mass scale \cite{ONEW}. In non-generic models of inflation the higher-order corrections may assume a specific form either because the inflaton has a particular symmetry or because the rate of inflaton roll remains constant (and possibly larger than $1$). Examples along this direction are certain fast-roll scenarios \cite{NON1,NON2,NON3} or the higher-order curvature corrections given in terms of the Gauss-Bonnet combination and weighted (in four space-time dimensions) by inflaton-dependent couplings\cite{NON4,NON5,NON6}. Similar modifications of the evolution of the tensor modes occurs in the case of Einstein-aether models \cite{NON7,NON8,NON9} or in the case of compact extra-dimensions \cite{DEC1a,DEC2a}. From the physical viewpoint a common aspect of different parametrizations is that the gravitational waves may acquire an effective index of refraction \cite{REFR1,REFR2}, as suggested long ago \cite{CC1a,CC1b} without any reference 
to the inflationary dynamics (see also \cite{PPNN}).   
Even if the geometry undergoes a stage of conventional accelerated expansion the intermediate slope of the spectral energy density increases depending on the evolution of the refractive index \cite{REFR1,REFR2}. 
The different contributions to the evolution of the tensor modes of the geometry 
in the case of conformally flat background geometries are summarized as follows:
\begin{equation}
S_{g} = \frac{1}{8 \ell_{P}^2} \int \, d^{3} x \int \, d \tau \biggl[ A(\tau) \partial_{\tau} h_{i\, j} \partial_{\tau} h^{i\, j} - B(\tau) \, \partial_{k} h_{i \, j} \partial^{k}h^{i\, j} 
- B_{c}(\tau) \overline{\gamma}^{A\, B} \partial_{A} h_{i \, j}  \partial_{B} h^{i \, j} \biggr].
\label{DEC1}
\end{equation}
In Eq. (\ref{DEC1})  the terms that would break parity and that are associated 
with quadratic combinations involving either dual Riemann tensor or the dual Weyl tensor 
have been neglected; both terms would appear in the effective action 
\cite{ONEW} (see also \cite{TWO}) and might in principle polarize the relic gravitons. Equation (\ref{DEC1}) contains three undetermined functions $A(\tau)$, $B(\tau)$ and $B_{c}(\tau)$. The coefficient $B(\tau)$ refers to the expanding dimensions while the presence of $B_{c}(\tau)$ 
is related to a possible mass term that arises from the internal (compact) dimensions
\footnote{In this situation the reduced Planck length is a function of the volume of the internal dimensions. We will generally work with the case of a spatially flat internal and external manifold with topology $M_{3+1}\times T_{d}$ where $(3+1)$ is the conventional $(3+1)$-dimensional flat Universe and $T_{d}$ is the $d$-dimensional torus. In what follows the Latin (lowercase) indices refer to the 3 (external) dimensions while the Latin (uppercase) indices refer to the internal dimensions.}.  In the case of a toroidal compactification (which is the one discussed here) $\overline{\gamma}_{A\, B} = \delta_{A\,B}$ \cite{DEC1a,DEC2a}. 
There are two equivalent ways in which Eq. (\ref{DEC1}) can be phrased in terms of an appropriate refractive index. The first possibility is to factor $A(\tau)$:
\begin{equation}
S_{g} = \frac{1}{8 \ell_{P}^2} \int \, d^{3} x \int \, d \tau A(\tau) \biggl[\partial_{\tau} h_{i\, j} \partial_{\tau} h^{i\, j} - \frac{1}{n^2(\tau)} \,\partial_{k} h_{i \, j} \partial^{k} h^{i\, j} - \frac{1}{n_{c}^2(\tau)} \overline{\gamma}^{A\, B} \partial_{A} h_{i \, j}  \partial_{B} h^{i \, j} \biggr],
\label{DEC2}
\end{equation}
where $n(\tau)$ and $n_{c}(\tau)$ denote the refractive indices associated, respectively, 
with the expanding and with the compact dimensions:
\begin{equation}
n(\tau) = \sqrt{A(\tau)/B(\tau)}, \qquad n_{c}(\tau) = \sqrt{A(\tau)/B_{c}(\tau)}.
\label{DEC3}
\end{equation}
The explicit form of the action (\ref{DEC3}) simplifies by changing 
the time parametrization and by rescaling the background dependence:
\begin{equation}
S_{g} = \frac{1}{8 \ell_{P}^2} \int \, d^{3} x \int \, d \eta \, \, c^2(\eta) \biggl[\partial_{\eta} h_{i\, j} \partial_{\eta} h^{i\, j} -  \, \, \partial_{k} h_{i \, j} \partial^{k} h^{i\, j} 
- r_{c}^2(\eta) \overline{\gamma}^{A\, B} \partial_{A} h_{i \, j}  \partial_{B} h^{i \, j} \biggr].
\label{DEC4}
\end{equation}
Equation (\ref{DEC4}) follows from Eq. (\ref{DEC2}) by changing the time 
parametrization and by redefining the background dependence according to:
\begin{equation}
n(\eta) \, d \eta = d \, \tau, \qquad c(\eta) = \sqrt{A(\eta)/n(\eta)}, \qquad r_{c}(\eta) = n(\eta)/n_{c}(\eta).
\label{DEC5}
\end{equation}
We could have made a different choice by factoring $B(\tau)$ instead of $A(\tau)$ in Eq. (\ref{DEC1}). This second choice is actually immaterial since the final result is exactly the same in the two cases. In fact if we first rescale $B(\tau)$ we simply get the analog of Eq. (\ref{DEC2}) which is:
\begin{equation}
S_{g} = \frac{1}{8 \ell_{P}^2} \int \, d^{3} x \int \, d \tau B(\tau) \biggl[ n^2(\tau)\, \partial_{\tau} h_{i\, j} \partial_{\tau} h^{i\, j} -  \, \partial_{k} h_{i \, j} \partial^{k} h^{i\, j} 
- \frac{B_{c}(\tau)}{B(\tau)}\, \overline{\gamma}^{A\, B} \partial_{A} h_{i \, j}  \partial_{B} h^{i \, j} \biggr].
\label{DEC6}
\end{equation}
If we again introduce the $\eta$-time defined as 
$n(\eta) \, d\eta = d\tau$, Eq. (\ref{DEC6}) becomes 
\begin{equation}
S_{g} = \frac{1}{8 \ell_{P}^2} \int \, d^{3} x \int \, d \eta \,\, \overline{c}^2(\eta) \,\,\biggl[ \, \partial_{\eta} h_{i\, j} \partial_{\eta} h^{i\, j} -  \, \partial_{k} h_{i \, j} \partial^{k} h^{i\, j} 
- r_{c}^2(\eta) \overline{\gamma}^{A\, B} \partial_{A} h_{i \, j}  \partial_{B} h^{i \, j} \biggr],
\label{DEC7}
\end{equation}
where $\overline{c}(\eta) = \sqrt{B(\eta)\, n(\eta)}$. By now comparing Eqs. (\ref{DEC4}) 
and (\ref{DEC7}) we see that $\overline{c}(\eta)$ and  $c(\eta)$ coincide since, in both cases,$\overline{c}(\eta) = c(\eta) = [ A(\eta)\, B(\eta)]^{1/4}$.

\subsection{Quantization and spectra in the $\eta$-time parametrization}
Since the two approaches mentioned above are equivalent we can introduce, as usual, the 
normal modes of the system $\mu_{i\, j}(\vec{x}, \eta) = c(\eta) \, h_{i\, j}(\vec{x}, \eta)$ so that the action takes the form:
\begin{eqnarray}
S_{g} &=& \frac{1}{8 \, \ell_{P}^2} \int d^{3} x\, \int d\eta\, \biggl[ \partial_{\eta} \mu_{i\, j} 
\partial_{\eta} \mu^{i\, j} + {\mathcal F}^2 \mu_{i\, j} \mu^{i\, j} -  {\mathcal F} \biggl( \mu_{i\, j} \partial_{\eta} \mu^{i\, j} + \mu^{i\, j} \partial_{\eta} \mu_{i\, j} \biggr) 
\nonumber\\
&-& \partial_{k} \mu_{i \, j} \partial^{k} \mu^{i\, j} - r^2(\eta) \overline{\gamma}^{A\, B} \partial_{A} \mu_{i\, j} \, \partial_{B} \mu^{i\, j}\biggr],
\label{calF0}
\end{eqnarray}
where ${\mathcal F}$ denotes the rate of variation of $c(\eta)$:
\begin{equation}
{\mathcal F} = \frac{\dot{c}}{c} = n \frac{c^{\,\prime}}{c},\qquad\qquad \dot{} = \partial_{\eta}, \qquad\qquad ^{\prime} = \partial_{\tau}.
\label{calF1}
\end{equation} 
In Eq. (\ref{calF1}) the overdot and the prime denote, respectively, a derivation with respect to $\eta$ and with respect to $\tau$. Three complementary time parametrizations are relevant for the present analysis and their features can be summarized, in short, as follows\footnote{Since the usual Hubble rate $H= \partial_{t} a/ a$ is defined in the cosmic time parametrization we remind that the overdot often denotes the derivative with respect to the cosmic time coordinate $t$ but, in the present context, the overdot will be reserved for the  derivation with respect to $\eta$-time, as suggested in Eq. (\ref{calF1}).}.
The $\eta$-time parametrization and the conformal time are 
related as $n(\eta) \, d\eta = d\tau$ (see also Eqs. (\ref{DEC2})--(\ref{DEC3})); the dictionary between the two is:
\begin{equation}
{\mathcal F} = \frac{\dot{c}}{c} = \frac{\partial \ln{c}}{\partial \eta} \equiv n\, a\,F,\qquad \qquad F = \frac{\partial \ln{c}}{\partial t}.
\label{calF2}
\end{equation}
The variation of the background geometry is typically expressed in terms of the cosmic 
time coordinate $t$ that is related to $\tau$ as $ a(\tau) d\tau = d t$ and, as usual, the connection between the rates of variation of the background is:
\begin{equation}
{\mathcal H} = \frac{a^{\prime}}{a} = \frac{\partial \ln{a}}{\partial \tau} \equiv a H,\qquad\qquad H = \frac{\partial \ln{a}}{\partial t}.
\label{calF2a}
\end{equation}
The rate of variation 
of the refractive index in units of the Hubble rate and the rate of variation of $H$ itself are then defined as:
\begin{equation}
\alpha = \frac{\partial \ln{n}}{\partial \ln{a}} = \frac{1}{n\, H} \frac{\partial n}{\partial t}, \qquad\qquad \epsilon = - \frac{\partial_{t}H}{H^2}\ll 1,
\label{calF3}
\end{equation}
where $\epsilon$ is the usual slow-roll parameter. Since the phase velocity coincides with the group velocity, the refractive index must increase during inflation (i.e. $\alpha \geq 0$) to prevent a superluminal propagation of the signal; there are, in practice, two relevant physical situations depending on the value of $\alpha$, i.e. $\alpha < 1$ and $\alpha = {\mathcal O}(1)$: in the first case $\alpha$ and $\epsilon$ are of the same 
order while in the second case $\alpha \gg \epsilon$. In what follows $\alpha$ is kept 
generic however, as we shall see, the tensor spectral index is determined by the competition of $\alpha$ and $\epsilon$ and the physical range corresponds to $\alpha < 1$.
We finally remark, as already mentioned, that the conventional 
slow-roll dynamics does not necessarily imply the validity of the so-called consistency relations which are instead broken by the presence of the refractive index so that,
 the tensor spectral index and the tensor-to-scalar ratio are not solely determined by $\epsilon$, as it happens when the consistency relations are enforced. 
 After these necessary specifications w can define the canonical momenta from Eq. (\ref{calF0})
\begin{equation}
\pi_{i\, j} = \frac{1}{8 \ell_{P}^2} \biggl[ \dot{\mu}_{i\, j} - {\mathcal F} \mu_{i\, j} \biggr], \qquad \pi^{i\, j} = \frac{1}{8 \ell_{P}^2} \biggl[ \dot{\mu}^{i\, j} - {\mathcal F} \mu^{i\, j} \biggr],
\label{calF4}
\end{equation}
so that the canonical Hamiltonian becomes:
\begin{eqnarray}
 H_{g}(\eta) &=& \int d^{3} x \biggl[ 8 \ell_{P}^2 \, \pi_{i\, j} \pi^{i\, j} +
{\mathcal F} \,\,\biggl(\mu_{i\, j} \pi^{i\, j} + \mu^{i\, j} \pi_{i\, j} \biggr)
\nonumber\\
&+& \frac{1}{8 \ell_{P}^2} \,\,\,\partial_{k} \mu_{i\, j} \partial^{k} \mu^{i\, j} + 
\frac{r^2(\eta)}{8 \ell_{P}^2} \,\,\,\overline{\gamma}^{A\, B} \partial_{A} \mu_{i\, j} \partial_{B} \mu^{i\, j} \biggr].
\label{calF5}
\end{eqnarray}
From Eq. (\ref{calF5}) the Hamilton's equations are:
\begin{equation}
\dot{\mu}_{i\, j} = 8 \ell_{P}^2 \,\pi_{i\, j} + {\mathcal F}\, \mu_{i\, j}, \qquad \qquad
\dot{\pi}_{i\, j} = - {\mathcal F} \, \pi_{i\, j} + \frac{\nabla^2 \mu_{i\, j}}{8 \ell_{P}^2} + 
r^2\, \frac{\overline{\nabla}^2 \mu_{i\, j}}{8 \ell_{P}^2},
\label{calF6}
\end{equation} 
where the Laplacian associated with the internal dimensions has been denoted by $\overline{\nabla}^2 = \overline{\gamma}^{A\, B}\partial_{A} \partial_{B}$. 
We can now quantize the system in the standard manner but, for the sake 
of accuracy, we repeat here the main steps. It is first useful to express the quantum field operators 
$\widehat{\mu}_{i\, j}(\vec{x},\eta)$ and $\widehat{\pi}_{m\, n}(\vec{x}, \eta)$ directly in Fourier space: 
\begin{equation}
\widehat{\mu}_{i\,j}(\vec{q}, \, \eta) = \frac{1}{(2\,\pi)^{3/2}} \int d^{3}x \,\,e^{i\, \vec{q}\cdot\vec{x}} \,\,\widehat{\mu}_{i\, j}(\vec{x},\eta), \qquad  \widehat{\pi}_{m\,n}(\vec{p}, \, \eta) = \frac{1}{(2\,\pi)^{3/2}} \int d^{3}x \,\,e^{i\, \vec{p}\cdot\vec{x}} \,\,\widehat{\pi}_{m\, n}(\vec{x},\eta).
\label{calF7}
\end{equation}
The field operators in Fourier space can then be expanded in the basis of the two 
(linear) tensor polarizations\footnote{The explicit form of the two linear polarizations is given by 
 $e_{ij}^{(\oplus)}(\hat{k}) = (\hat{m}_{i} \hat{m}_{j} - \hat{n}_{i} \hat{n}_{j})$ and by
$ e_{ij}^{(\otimes)}(\hat{k}) = (\hat{m}_{i} \hat{n}_{j} + \hat{n}_{i} \hat{m}_{j})$
where $\hat{k}_{i} = k_{i}/|\vec{k}|$,  $\hat{m}_{i}$ and $\hat{n}_{i}$ are three mutually 
orthogonal unit vectors obeying $\hat{m}\times \hat{n} = \hat{k}$.}:
\begin{equation}
\widehat{\mu}_{i\,j}(\vec{q}, \, \eta) = \sum_{\lambda= \oplus,\,\otimes} \,\, e^{(\lambda)}_{i\, j}(\hat{q})  \,\, \widehat{\mu}_{\lambda}(q, \eta), \qquad 
\widehat{\pi}_{m\,n}(\vec{p}, \, \eta) = \sum_{\lambda= \oplus,\,\otimes} \,\, e^{(\lambda)}_{m\, n}(\hat{p}) \,\, \widehat{\pi}_{\lambda}(p, \eta).
\label{calF8}
\end{equation}
The field operators of Eq. (\ref{calF8}) can be directly written in terms of the 
creation and annihilation operators obeying $[\widehat{a}_{\vec{q}, \, \lambda}, \, \widehat{a}^{\dagger}_{\vec{p},\, \lambda^{\prime}}] = \delta^{(3)}(\vec{q}- \vec{p})$ so that, ultimately, $\widehat{\mu}_{i\,j}(\vec{q}, \, \eta)$
and $\widehat{\pi}_{m\,n}(\vec{p}, \, \eta)$ are:
\begin{eqnarray}
\widehat{\mu}_{i\, j}(\vec{q}, \eta) &=&  \sqrt{2} \ell_{P} \, \sum_{\lambda} \biggl[ e_{i\,j}^{(\lambda)}(\hat{q}) f_{k,\, \lambda}(\eta) \hat{a}_{\vec{q}\, \lambda} +
e_{i\,j}^{(\lambda)}(-\hat{q})  f_{k,\, \lambda}^{*}(\eta)  \hat{a}_{-\vec{q}\, \lambda}^{\dagger} \biggr],
\label{calF9}\\
\widehat{\pi}_{m\, n}(\vec{p}, \eta) &=&  \frac{1}{4 \, \sqrt{2} \, \ell_{P}}, \sum_{\lambda} \biggl[ e_{m\,n}^{(\lambda)}(\hat{p}) g_{k,\, \lambda}(\eta) \hat{a}_{\vec{p}\, \lambda} +
e_{m\,n}^{(\lambda)}(-\hat{p})  g_{k,\, \lambda}^{*}(\eta)  \hat{a}_{-\vec{p}\, \lambda}^{\dagger} \biggr].
\label{calF10}
\end{eqnarray}
It can be directly checked that the commutation relations between $\widehat{\mu}_{i\, j}(\vec{q}, \eta)$ and $\widehat{\pi}_{m\, n}(\vec{p}, \eta)$ are given by:
\begin{equation}
[ \widehat{\mu}_{i\, j}(\vec{q}, \eta),\, \widehat{\pi}_{m\, n}(\vec{p}, \eta) ] = i\, {\mathcal S}_{i\, j\, m\, n}(\hat{q}) \delta^{(3)}(\vec{q} + \vec{p}),
\label{calF11}
\end{equation}
where ${\mathcal S}_{i\, j\, m\, n}(\hat{q})$
\begin{equation}
{\mathcal S}_{i\,j\,m\,n}(\hat{q}) = \frac{1}{4} \biggl[ p_{mi}(\hat{q}) p_{nj}(\hat{q}) + p_{mj}(\hat{q}) p_{ni}(\hat{q}) - p_{ij}(\hat{q}) p_{m n}(\hat{q}) \biggr],
\label{calF12}
\end{equation}
and $p_{i\, j} = (\delta_{i\,j} - \hat{q}_{i}\, \hat{q}_{j})$. Note that Eq. (\ref{calF11}) 
holds provided the Wronskian normalization condition is verified:
\begin{equation}
f_{q,\, \lambda}(\eta) g^{*}_{q,\, \lambda}(\eta) - f_{q,\, \lambda}^{*}(\eta) g_{q,\, \lambda}(\eta) = i.
\label{calF13}
\end{equation}
It is finally practical to deduce the two-point functions of $\hat{h}_{ij}(\vec{k},\eta)$ and of 
$\partial_{\eta} \hat{h}_{ij}(\vec{k},\eta)$ in Fourier space: 
\begin{eqnarray}
\langle \hat{h}_{ij}(\vec{k},\eta) \, \hat{h}_{mn}(\vec{p},\eta) \rangle &=& \frac{2\pi^2}{k^3} P_{T}(k,\eta) \, {\mathcal S}_{i\,j\,m\,n}(\hat{k}) \delta^{(3)}(\vec{k} +\vec{p}),
\label{calF14}\\
\langle \partial_{\eta} \hat{h}_{ij}(\vec{k},\eta) \,\, \partial_{\eta}\hat{h}_{mn}(\vec{p},\eta) \rangle &=& \frac{2\pi^2}{k^3} Q_{T}(k,\eta) \, {\mathcal S}_{i\,j\,m\,n}(\hat{k}) \delta^{(3)}(\vec{k} +\vec{p}),
\label{calF15}
\end{eqnarray}
where the power spectra $P_{T}(k,\tau)$ and $Q_{T}(k,\tau)$ are defined, respectively, by: 
\begin{eqnarray}
P_{T}(k,\eta) = \frac{4 \ell_{P}^2\, \,k^3}{\pi^2 c^2(\eta)} |f_{k}(\eta)|^2, \qquad Q_{T}(k,\eta) = \frac{4 \ell_{P}^2\,\, k^3}{\pi^2 c^2(\eta)} |g_{k}(\eta)|^2.
\label{calF17}
\end{eqnarray}
All in all, putting everything together, we have that the field operators are expressed as:
\begin{eqnarray}
\widehat{\mu}_{ij}(\vec{x},\eta) &=& \frac{\sqrt{2} \ell_{P}}{(2\pi)^{3/2} }\sum_{\lambda=\oplus,\,\otimes} \int \, d^{3} k \,\,e^{(\lambda)}_{ij}(\vec{k})\, \biggl[ f_{k,\lambda}(\eta) \hat{a}_{\vec{k}\,\lambda } e^{- i \vec{k} \cdot \vec{x}} + f^{*}_{k,\lambda}(\eta) \hat{a}_{\vec{k}\,\lambda }^{\dagger} e^{ i \vec{k} \cdot \vec{x}} \biggr],
\label{calF18}\\
\widehat{\pi}_{ij}(\vec{x},\eta) &=& \frac{1}{4 \, \sqrt{2} \, \ell_{P} \,(2\pi)^{3/2} }\sum_{\lambda=\oplus,\,\otimes} \int \, d^{3} k \,\,e^{(\lambda)}_{ij}(\vec{k})\, \biggl[ g_{k,\lambda}(\eta) \hat{a}_{\vec{k}\,\lambda } e^{- i \vec{k} \cdot \vec{x}} + g^{*}_{k,\lambda}(\eta) \hat{a}_{\vec{k}\,\lambda }^{\dagger} e^{ i \vec{k} \cdot \vec{x}} \biggr],
\label{calF19}
\end{eqnarray}
where, according to Eq. (\ref{calF6}),  the evolution of the mode functions $f_{k,\lambda}$ and $g_{k,\lambda}$ obeys:
\begin{eqnarray}
\dot{f}_{k,\,\lambda} &=& g_{k,\, \lambda} + {\mathcal F} \, f_{k,\,\lambda},
\label{NM12a}\\
\dot{g}_{k,\,\lambda} &=& - k^2 f_{k,\,\lambda} -  {\mathcal F}  f_{k,\,\lambda}  - q^2 \, r^2\, f_{k,\lambda}.
\label{NM12b}
\end{eqnarray}
Thanks to the coupling among the scalar and the tensor modes the gravity wave evolution equation get what looks like a massive contribution \cite{DEC1a,DEC2a}. In terms of the eigenstates of the Laplace operators appearing in Eq. (\ref{calF6}) we have that $\nabla^2 \mu_{i\,j}= - k^2 \mu_{i\,j}$ and $\overline{\nabla}^2 \mu_{i\,j}= - q^2 \mu_{i\,j}$.
While $k$ denotes the external momentum, $q$ is the momentum associated with the extra-dimensions which can be viewed as a massive 
contribution as it can be appreciated by decoupling Eqs. (\ref{NM12a})--(\ref{NM12b}):
\begin{equation}
\ddot{f}_{k}  + \biggl[k^2 + q^2 r^2 - \frac{\ddot{c}}{c} \biggr] f_{k} =0, \qquad g_{k}= \dot{f}_{k} - {\mathcal F} \, f_{k};
\label{NM13}
\end{equation}
the polarization index has been omitted since the result of Eq. (\ref{NM14}) 
holds both for $\oplus$ and for $\otimes$. 

\subsection{Different physical limits} 
Depending on the values of $c(\eta)$, $n(\eta)$ and $r(\eta)$, the action of Eq. (\ref{calF0})  describes a number of relevant situations that are however physically different. In the standard limit the refractive index is absent and the internal dimensions disappear: \begin{equation}
n\to 1,\qquad\qquad r\to 0,\qquad\qquad \eta \to \tau, \qquad\qquad c(\tau) = a(\tau),
\label{LIMIT1}
\end{equation}
In the limit (\ref{LIMIT1}) the rescaled normal mode becomes $\mu_{i\, j} = a(\tau) h_{i\, j}$
and $a(\tau)$ is the scale factor appearing in the four-dimensional line 
element 
\begin{equation}
d s^2 = \overline{g}_{\mu\nu} \, d\,x^{\mu}\,\, x \, x^{\nu}, \qquad\qquad \overline{g}_{\mu\nu} = a^2(\tau) \eta_{\mu\nu},
\label{LIMIT1a}
\end{equation}
where $\eta_{\mu\nu}$ is the Minkowski metric. 
According to Eq.  Eq. (\ref{calF0}) coincides with the original Ford-Parker action \cite{par1}; furthermore, as 
stressed in Eq. (\ref{LIMIT1}), the $\eta$-time and the conformal time 
coordinates coincide. In the case of Eq. (\ref{LIMIT1}) the only possibility 
of getting a flat spectrum at high-frequency is to modify the standard inflationary 
dynamics by considering, for instance, the possibility of bouncing backgrounds.
In this case Eq. (\ref{NM13}) becomes:
\begin{equation}
f_{k}^{\prime\prime}  + \biggl[k^2  - \frac{a^{\prime\prime}}{a} \biggr] f_{k} =0, \qquad g_{k}= f_{k}^{\prime} - {\mathcal H} \, f_{k}.
\label{NM14}
\end{equation}
Let us now suppose that $n\to 1$ in the presence of  $d$ internal dimensions characterized by the scale 
factor $b^2(\tau)$ so that the line element  is given by:
\begin{equation}
ds^2 =  a^2(\tau)  [ d \tau^2 \,-\, d\vec{x}^2] - b^2(\tau) \overline{\gamma}^{A\, B} \,d y_{A} \, d y_{B},
\label{INTMET}
\end{equation}
where $A,\, B= 1, .\,.\,.\,, d$ runs over the $d$ internal dimensions and the 
dimensionality of the space-time is $D= 4 + d$. As already suggested, we mainly consider the case $\overline{\gamma}_{A\, B} = \delta_{A\, B}$. In the 
case of Eq. (\ref{INTMET}) the internal volume is $b^{d/2}$ and we then have that the various 
parameters 
\begin{equation}
 n\to 1,\qquad\qquad r\to \frac{a(\tau)}{b(\tau)},\qquad\qquad \eta \to \tau, \qquad\qquad c(\tau) = a(\tau) \, b^{d/2}(\tau).
\label{LIMIT2}
\end{equation}
In the situation of Eq. (\ref{LIMIT2}) the conformal time coincides with the $\eta$-time but the presence of $b(\tau)$ accounts 
for the dynamics of the $d$-extra-dimensions. According to Eq. (\ref{LIMIT2}) the explicit form of Eq. (\ref{NM13}) becomes 
\begin{equation}
f_{k}^{\prime\prime}  + \biggl[k^2  + q^2 \frac{a^2}{b^2}- \frac{(a\, b^{d/2})^{\prime\prime}}{a \, b^{d/2}} \biggr] f_{k} =0, \qquad g_{k}= f_{k}^{\prime} - {\mathcal F} \, f_{k} , \qquad {\mathcal F} = {\mathcal H} + \frac{d}{2} \frac{b^{\prime}}{b}.
\label{NM15}
\end{equation}
A scenario of dimensional decoupling based on Eq. (\ref{NM15}) has been discussed in Refs. \cite{DEC1a,DEC2a} and the considerations reported here can be easily extended to that situation. We finally consider the framework that is more realistic, at least for the present ends:\begin{equation}
r \to 0, \qquad n(\eta) d\eta = d\tau, \qquad c(\eta) = \frac{a(\eta)}{\sqrt{n(\eta)}}.
\label{LIMIT3}
\end{equation}
In the limit (\ref{LIMIT3})  the evolution of the mode functions (see Eq. (\ref{NM13})) becomes:
\begin{equation}
\ddot{f}_{k}  + \biggl[k^2  - \frac{\ddot{c}}{c} \biggr] f_{k} =0, \qquad g_{k}= \dot{f}_{k} - {\mathcal F} \, f_{k}.
\label{NM16}
\end{equation}
 In the present discussion we 
also assume that the refractive index is exactly $1$ at the present time. It is useful
to remark that the evolution of the refractive index is specified unambiguously by assigning $n(a)$.  Even though the phase velocity of the relic gravitons is not required to be sub-luminal we consider here the situation where $n(a) \geq 1$. When $n(a)$ changes appreciably during inflation and it goes to $1$ in the standard decelerated stage of expansion\footnote{In Eq. (\ref{NEX}) $a_{i}$ and $a_{1}$ mark, respectively, the beginning and the end of the inflationary epoch; $a_{*}$ defines the boundary of the refractive stage and $N_{*}$ is the corresponding number of $e$-folds.}:
\begin{equation}
n(a) = n_{\ast} \frac{ (a/a_{\ast})^{\alpha} \,\,e^{- \gamma (a/a_{1})}}{(a/a_{*})^{\alpha} + 1} + 1, \qquad\qquad
n_{\ast} = n_{i} (a_{\ast}/a_{i})^{\alpha} = n_{i} e^{\alpha \, N_{\ast}}.
\label{NEX}
\end{equation}
Equation (\ref{NEX}) defines, in practice, three successive physical regimes. For $a \gg a_{1}$ the refractive index goes to $1$ and the standard situation is recovered
depending on the value of $\gamma\geq 1$ which controls the sharpness of the transition.
When $a_{*} < a < a_{1}$ the refractive index is practically constant but still larger than $1$, i.e. $n(a)\simeq n_{\ast} > 1$. Finally for $a< a_{\ast}$ we have the truly refractive stage where $n (a) \simeq n_{\ast} (a/a_{\ast})^{\alpha}$.

\subsection{Chirp amplitude, spectral amplitude and spectral energy density}
When discussing the relic gravitons at the present time the observational collaborations  typically assume that the space-time is flat and that the frequency of the gravitons 
is always larger than the rate of variation of the geometry which means, in terms 
of the previous notations, that $k\eta \gg 1$.
 When we are in flat space-time and under 
the conditions of Eq. (\ref{NEX}) we have that the conformal, the cosmic and the 
$\eta$-time all coincide at the present epoch i.e.
\begin{equation}
\eta = \tau = t , \qquad\qquad c_{0} = a_{0} =1.
\label{NEX0}
\end{equation}
Furthermore, since the scale factor is normalized to $1$, the comoving 
and the physical frequencies are (today) concident. 
To introduce the spectral amplitude the simplest 
approach is to expand the tensor amplitude as 
\begin{equation}
h_{i\, j}(\vec{x}, \eta) = \int_{-\infty}^{\infty} d \nu \int d\,\widehat{k}  \, e^{ 2\,i\,\pi \, \nu\,( \eta - \widehat{k}\cdot\vec{x})} \,h_{i\, j}(\nu, \widehat{k}), \qquad 
\,h^{*}_{i\, j}(\nu, \widehat{k}) = \,h_{i\, j}(-\nu, \widehat{k}),
\label{NEX1}
\end{equation}
where $\nu = k/(2\pi)$ is the comoving frequency and $d\widehat{k} = d\cos{\vartheta} \, d\varphi$. As before $h_{i\, j}(\nu, \widehat{k})$ can be expanded in the basis of the linear  polarisations $\oplus$ and $\otimes$:
\begin{equation}
h_{i\, j}(\nu, \widehat{k}) = \sum_{\lambda= \oplus, \, \otimes} \, e_{i\, j}^{(\lambda)}(\widehat{k}) \, h_{\lambda}(\nu, \widehat{k}). 
\label{NEX2}
\end{equation}
The spectral amplitude $S_{h}(|\nu|)$ is defined as the expectation value of the tensor amplitudes 
expressed as a function of $\nu$ and $\widehat{k}$:
\begin{equation}
\langle h_{\lambda} (\nu, \, \widehat{k}) \, h_{\lambda^{\prime}} (\nu^{\prime}, \, \widehat{k}^{\prime}) \rangle = 
{\mathcal A}_{c}\,\, S_{h}(|\nu|) \,\, \delta(\nu + \nu^{\prime})\, \, \delta^{(2)}(\widehat{k} - \widehat{k}^{\,\prime}) \,\, \delta_{\lambda\,\lambda^{\prime}};
\label{OBS13a}
\end{equation}
where ${\mathcal A}_{c}$ is an overall constant that parametrizes the different choices 
currently adopted by different authors\footnote{In the definition of the spectral amplitude 
we introduced a modulus since we intend to express the various integrations 
for positive values of $\nu$.}. Consistently with the previous notations, the angular delta function appearing in Eq. (\ref{OBS13a}) is given by $\delta^{(2)}(\widehat{k} - \widehat{k}^{\,\prime}) = \delta(\varphi -\varphi^{\prime})\, \delta(\cos{\vartheta} - \cos{\vartheta}^{\prime})$.
If we now compute the expectation value of two tensor amplitudes with different indices 
we obtain
\begin{equation}
\langle h_{i\, j}(\nu, \widehat{k}) h_{\ell\, m}(\nu^{\prime}, \widehat{k}^{\prime}) \rangle = 4 \, {\mathcal A}_{c}\,{\mathcal S}_{i\, j\, \ell\, m}(\widehat{k}) \, S_{h}(|\nu|) \, \delta^{(2)}(\widehat{k} - \widehat{k}^{\prime}) \, \delta(\nu + \nu^{\prime}),
\label{OBS13c}
\end{equation}
where ${\mathcal S}_{i\, j\, \ell\, m}(\widehat{k})$ has been already introduced in Eq. (\ref{calF12}). The expectation value of the tensor amplitudes 
at equal time becomes:
\begin{eqnarray}
\langle h_{i\,j}(\vec{x}, \eta)\, h^{i \, j}(\vec{y},\eta) \rangle &=& \int_{-\infty}^{+\infty} d \, \nu \,
\int_{-\infty}^{+\infty} d \, \nu^{\prime}\, 
\int \, d\widehat{k} \,  \int \, d\widehat{k}^{\prime}
\nonumber\\
&\times& e^{ 2 i \pi \, \nu(\tau - \widehat{k}\cdot\vec{x})} \,\,  e^{ 2 i \pi \, \nu(\tau - \widehat{k}^{\prime}\cdot\vec{y})}
\langle \, h_{i\, j} ( \nu, \widehat{k}) \, \, h^{i\, j} ( \nu^{\prime}, \widehat{k}^{\prime}) \rangle.
\label{REDEFS2}
\end{eqnarray}
The expectation value appearing in Eq. (\ref{REDEFS2}) can be directly 
computed thanks to Eqs. (\ref{OBS13c}). More specifically, 
since ${\mathcal S}_{i\,j\,i\,j} =1$ Eq. (\ref{REDEFS2}) becomes:
\begin{equation}
\langle h_{i\,j}(\vec{x}, \eta)\, h^{i \, j}(\vec{y},\eta) \rangle = 32 \pi {\mathcal A}_{c} 
\int_{0}^{\infty} d \nu \, S_{h}(|\nu|) \, j_{0}(2 \pi \,\nu\,r).
\label{REDEFS3}
\end{equation}
From the direct comparison of Eq. (\ref{REDEFS3}) with the analog expression 
computed in terms of $P_{T}(\nu)$ it follows that the relations between the chirp amplitude, the spectral 
amplitude and the power spectrum is:
\begin{equation}
h_{c}^2(\nu) = 16 \pi {\mathcal A}_{c} \, \nu S_{h}(|\nu|),\qquad\qquad
P_{T}(\nu)  = 32 \pi {\mathcal A}_{c} \, \nu S_{h}(|\nu|).
\label{REDEFS4}
\end{equation}
The specific values of ${\mathcal A}_{c}$ can be used to rationalize the 
obtained expressions. The LIGO/Virgo collaboration is normally setting ${\mathcal A}_{c} = 1/(16\, \pi)$ so that Eq. (\ref{REDEFS4}) becomes 
\begin{equation}
h_{c}^2(\nu) =   \nu S_{h}(|\nu|),\qquad
P_{T}(\nu)  = 2 \, \nu S_{h}(|\nu|), \qquad \langle h_{i\,j}(\vec{x}, \tau)\, h^{i \, j}(\vec{x},\tau) \rangle = 2
\int_{0}^{\infty} d \nu \, S_{h}(|\nu|).
\label{REDEFS5}
\end{equation}
The PTA collaborations express their results in terms of the chirp amplitude $h_{c}^2(\nu)$. In this respect we just note that, up to a numerical factor, the square of the chirp amplitude coincides with the power spectrum so that its relation with the spectral energy density may be easily determined and it is:
\begin{equation} 
P_{T}(\nu, \tau_{0}) = 2 \, h_{c}^2(\nu,\tau_{0}), \qquad \qquad
\Omega_{gw}(\nu, \tau_{0}) = \frac{2 \pi^2}{3 H_{0}^2} \, \nu^2 \, h_{c}^2(\nu, \tau_{0}).
\label{NOTT0}
\end{equation}

\renewcommand{\theequation}{3.\arabic{equation}}
\setcounter{equation}{0}
\section{Normalization in the audio band and in the nHz domain}
\label{sec3}
\subsection{Wide-band interferometers and the audio band}
The observations of wide-band detectors led through the years to a number direct upper
limits on the backgrounds of relic gravitons for a frequency interval ranging between few Hz and $10$ kHz \cite{STone,STonea,STtwo,STthree,STthreea,STfour,STfive,STsix}. 
The most relevant upper limits are summarized in Tab. \ref{TABLE1} and they depend on the spectral slope of the signal.
\begin{table}[!ht]
\begin{center}
\caption{List of the direct limits on the relic gravitons obtained by wide-band interferometers.}
\vskip 0.4 cm
\begin{tabular}{||c|c|c|c||}
\hline
\hline
\rule{0pt}{4ex}  Year & frequency range [Hz] & Bound & Reference \\
\hline
\hline
& & &\\
$2004$ & $40-314$ & $ \overline{\Omega}(0) < 23$ & Ref. \cite{STone} \\
$2005$ & $69-156$ & $  \overline{\Omega}(0) < 8.4 \times 10^{-4}$  & Ref. \cite{STonea} \\
$2012$ & $600-1000$ & $\overline{\Omega}(3) < 0.32$ & Ref. \cite{STtwo} \\
$2014$ & $41.5-169.25$ & $\overline{\Omega}(0) < 5.6 \times 10^{-6}$ & Ref. \cite{STthree}\\
$2014$ & $600-1000$ & $\overline{\Omega}(3) < 0.14$ & Ref. \cite{STthree}\\
$2014$ & $170-600$ & $\overline{\Omega}(0) < 1.8\times 10^{-4}$ & Ref. \cite{STthree}\\
$2014$ & $1000-1726$ & $\overline{\Omega}(3) < 1$ & Ref. \cite{STthree}\\
$2015$ & $460-1000$  & $\overline{\Omega}(3) < 7.7 \times 10^{-4}$ & Ref. \cite{STthreea}\\
$2017$ & $20-86$ &    $\overline{\Omega}(0)< 1.7 \times 10^{-7}$ & Ref. \cite{STfour}\\
$2017$ & $20-300$ & $\overline{\Omega}(3) < 1.7 \times 10^{-8}$ & Ref. \cite{STfour}\\
$2019$  & $20-81.9$ & $\overline{\Omega}(0) < 6 \times 10^{-8}$  & Ref. \cite{STfive}\\
$2019$  & $20-95.2$ & $\overline{\Omega}(2/3) < 4.8 \times 10^{-8}$ &Ref. \cite{STfive}\\
$2019$ &  $20-301$  & $\overline{\Omega}(3) < 7.9 \times 10^{-9}$ & Ref. \cite{STfive}\\
$2021$ &  $20-76.6$ & $\overline{\Omega}(0) < 5.8\times 10^{-9}$ & Ref. \cite{STsix}\\
$2021$ &  $20-90.6$ & $\overline{\Omega}(2/3) < 3.4\times 10^{-9}$ & Ref. \cite{STsix}\\
$2021$ & $ 20-291.6$ & $\overline{\Omega}(3) < 3.9\times 10^{-10}$ & Ref. \cite{STsix}\\ 
\hline
\hline
\end{tabular}
\label{TABLE1}
\end{center}
\end{table}
For the purposes of Tab. \ref{TABLE1} the spectral energy density has been parametrized with a power-law slope of the type: 
\begin{equation}
\Omega_{gw}(\nu) = \overline{\Omega}(\sigma) \, \biggl(\frac{\nu}{\nu_{ref}} \biggr)^{\sigma}, \qquad \sigma \geq 0,
\label{SNR2}
\end{equation}
where $\nu_{ref}$ is a (conventional) frequency while $\overline{\Omega}(\sigma)$ is the constant amplitude that differs\footnote{For instance, $\overline{\Omega}(0)$ is the amplitude of the scale-invariant spectral energy density while $\overline{\Omega}(3)$ is the amplitude of a spectral energy density with cubic slope. } depending on the slope $\sigma$. The results of Tab. \ref{TABLE1} show that the constant amplitudes associated with the various $\sigma$ are constrained at different levels. The scale-invariant limit represents the simplest signal grossly compatible with the concordance paradigm; conversely,  if $\sigma= 3$ in Eq. (\ref{SNR2}),  the factor $\Omega_{gw}(\nu)^2/\nu^{6}$ is practically constant and, in this case, the estimate of the integral appearing in the signal-to-noise ratio gets simpler (see e.g. \cite{SN1,SN2,SN3,SN4}). In Refs.\cite{STone,STonea,STtwo,STthree,STthreea,STfour} the LIGO/Virgo collaboration  presented the upper limits for $\Omega(0)$ and $\Omega(3)$; a third case has been subsequently analyzed and it corresponds to $\sigma=2/3$ \cite{STfive,STsix}. In the case of an exactly scale-invariant spectrum the constraints obtained in Refs. \cite{STfour,STfive} imply that $\overline{\Omega}(0) < 6\times 10^{-8}$. If this value is compared with the analog result obtained in Ref. \cite{STthree}, the upper limit is ${\mathcal O}(100)$ times more constraining for the same slope and for the same frequency band.   
For $\sigma = 2/3$ the constraints of Ref. \cite{STfour} imply $\overline{\Omega}_{2/3} < 4.8 \times 10^{-8}$ with 95\% confidence within the $20$--$95$ Hz frequency band with $\nu_{ref} = 25$ Hz.  The slope $\sigma= 2/3$ may actually parametrize a potentially interesting foreground for the 
relic gravitons\footnote{Depending on the estimates, the rates of black hole  mergers range from ${\mathcal O}(50) \, \mathrm{Gpc}^{-3}\, \mathrm{yr}^{-1}$ to ${\mathcal O}(300) \, \mathrm{Gpc}^{-3}\, \mathrm{yr}^{-1}$. If this is the case we can not only expect to have many more signals 
but also a stochastic foreground coming from  unresolved sources of gravitational radiation.}.
The most constraining limit to date in the case $\sigma =2/3$ has been obtained 
in Ref. \cite{STsix} by the Kagra-Ligo-Virgo collaboration and it requires that 
$\overline{\Omega}(2/3) < 3.4\times 10^{-9}$. The most constraining set of bounds appearing in Tab. \ref{TABLE1} corresponds to the one obtained by the LIGO, Virgo and KAGRA collaborations \cite{STsix}. In the case of a flat spectral energy density the bound reads\footnote{In what follows, 
for the sake of conciseness, when the time dependence is suppressed it is understood 
that the corresponding quantity is evaluated at the present time. So, for instance, 
$\Omega_{gw}(\nu)= \Omega_{gw}(\nu, \tau_{0})$, $h_{c}(\nu,\tau_{0})= h_{c}(\nu)$ and so on and so forth.}: 
\begin{equation}
\Omega_{gw}(\nu_{L}) < 5.8 \times 10^{-9}, \qquad\qquad 20 \,\, \mathrm{Hz} < \nu_{L} < 76.6 \,\, \mathrm{Hz},
\label{CONS2}
\end{equation}
where $\nu_{L}$ denotes the LIGO-Virgo-KAGRA frequency; we shall commonly refer to this limit as the LVK bound. Even if in Eq. (\ref{CONS2}) we just quoted the most constraining limit, the LIGO-Virgo-KAGRA  collaboration actually reports a threefold bound for three different values of $\sigma$. When the value of $\sigma$ increases the bound becomes more restrictive 
once the reference frequency is kept fixed. The three results are unified in the following interpolating formula
\begin{equation}
\log{\overline{\Omega}}(\sigma) < -\,8.236 -\, 0.335\, \sigma- 0.018\, \sigma^2.
\label{NOT2}
\end{equation}
that will be used to impose the LVK bound at high-frequency.
\subsection{Pulsar timing arrays and the nHz band}
Equations (\ref{CONS2})--(\ref{NOT2}) are just the last result of the series of bounds given 
in Tab. \ref{TABLE1} and it is not excluded they might be further improved 
in the near future. Already at the present stage, however, they are quite interesting 
if they are combined with the current evidences provided by the Pulsar Timing 
Arrays (PTA). 
\begin{table}[!ht]
\begin{center}
\caption{List of current measurements from the various Pulsar Timing Arrays. The typical 
reference frequency is taken to be $\nu_{ref} = 31.68\,\mathrm{nHz}$.}
\vskip 0.4 cm
\begin{tabular}{||c|c|c|c|c||}
\hline
\hline
\rule{0pt}{4ex} Experiment & $q$ & $\beta$ & $h_{0}^2 \, \Omega_{gw}(\nu_{ref},\tau_{0})$ & Reference \\
\hline
\hline
& & & &\\
Nanograv & $1.92$ & $ -2/3$ & $2.31\times 10^{-9}$ &Ref. \cite{NANO1} \\
PPTA & $2.2$ & $  -2/3$ & $3.04 \times10^{-9}$ & Ref. \cite{PPTA} \\
EPTA & $2.95$ & $-2/3$ & $5.47\times 10^{-9}$ & Ref. \cite{EPTA} \\
EPTA & $5.13$ & $-0.33$ & $1.65\times 10^{-8}$ & Ref. \cite{EPTA}\\
IPTA & $2.8$ & $-2/3$ & $4.93\times 10^{-9}$ & Ref. \cite{IPTA}\\
IPTA & $3.8$ & $-0.5$ & $9.08\times 10^{-9}$ & Ref. \cite{IPTA}\\
\hline
\hline
\end{tabular}
\label{TABLE2}
\end{center}
\end{table}
In Tab. \ref{TABLE2} we illustrate the main findings of the four different PTA collaborations \cite{NANO1,PPTA,EPTA,IPTA}. Before 
going through the details we can already see that the determinations of the spectral 
energy density seem to be, naively, at the same level of the LVK bound but in a 
different frequency domain. The two results of Ref. \cite{EPTA} differ because 
of the different values of $\beta$ and the same comment also holds in the case 
of Ref. \cite{IPTA}.  The various PTA collaborations \cite{NANO1,PPTA,EPTA,IPTA} express the chirp amplitude 
at a pivot frequency $\nu_{ref} = 31.68\,\, \mathrm{nHz}$ corresponding to $\mathrm{yr}^{-1}$:
\begin{equation}
h_{c}(\nu) = \,Q \,\biggl(\frac{\nu}{\nu_{ref}}\biggr)^{\beta}, \qquad \qquad \nu_{ref} = \frac{1}{\mathrm{yr}}=   31.68\,\, \mathrm{nHz}.
\label{NOTT1}
\end{equation}
The pivotal model analyzed so far assumes $\beta = -2/3$ and this is the case preferentially 
reported in Tab. \ref{TABLE2}. The EPTA \cite{EPTA} and IPTA \cite{IPTA} 
collaborations also consider a more general class of scenarios where $Q$ and $\beta$ 
vary simultaneously. For instance 
the EPTA finds that the most favoured model to be the common uncorrelated red noise 
described by ${\mathcal Q}= 5.13^{+4.20}_{-2.73}\times 10^{-15}$ with $\overline{\gamma} = 3.78^{+0.69}_{-0.59}$ where we recall that, within the present notations, $\beta = (3-\overline{\gamma})/2$. If the spectral index is instead fixed as $\overline{\gamma} = 13/3$ (i.e. $\beta = -2/3$) Ref. \cite{EPTA} suggests  ${\mathcal Q}= 2.95^{+0.89}_{-0.72}\times 10^{-15}$. We 
not that $\overline{\gamma}$ will is not employed hereunder and are it is only mentioned for the sake of accuracy 
since some of the PTA collaborations introduce this notation which is a bit contrived from the viewpoint of the present discussion.

For the different estimates of Refs. \cite{NANO1,PPTA,EPTA,IPTA} the value of $Q$ 
is always ${\mathcal O}(10^{-15})$ it is therefore useful to express $Q = q_{0} \times 10^{-15}$.  
Using this notation and Eq. (\ref{NOTT1})  the spectral energy density in the nHz band can be finally expressed as\footnote{ For instance the PPTA collaboration \cite{PPTA}  suggests $q_{0}= 2.2$; the IPTA estimates $q_{0}= 2.8$ \cite{IPTA} while the EPTA \cite{EPTA} gives $q_{0} = 2.95$.  The results of PPTA, IPTA and EPTA seem, at the moment, to be broadly compatble with the NANOgrav 12.5 yrs data \cite{NANO1} implying $q_{0} =1.92$. }:
\begin{equation}
h_{0}^2\,\Omega_{gw}(\nu) = 6.290\times 10^{-10} \,\, \, q_{0}^2\, \biggl(\frac{\nu}{\nu_{ref}}\biggr)^{ 2 + 2 \beta}.
\label{NOTT8}
\end{equation}
The potential signal of the PTA  recently reported evidence 
of a potential signal in the nHz band. Using the spectral energy density in critical units as a pivotal variable the features of this purported signal would imply, in the present notations, that:
\begin{equation}
10^{-9.88}  \,\,< \frac{\, h_{0}^2\,\Omega_{gw}(\nu_{P})}{q_{0}^2} < \,\,10^{-8.86} , \qquad\qquad 3\,\,\mathrm{nHz} \, < \nu_{P}< \,100 \,\, \mathrm{nHz}.
\label{PTAb1}
\end{equation}

\subsection{Audio band and nHz band: which is the most restrictive?}
To avoid potential confusions it is relevant to compare the limits in the audio band and in the nHz to 
understand which are the most restrictive. Before discussing this comparison it is equally important to stress 
that the two classes of measurements are qualitatively different: while the limits from the audio band
are specific to the case of relic gravitons \cite{STfive,STsix}, the property of a PTA is that the signal 
from relic gravitons should be correlated across the baselines while that from the other noise will not. 
Since these correlation have not been observed so far, the interpretation suggested in 
by the PTA is still preliminary and the measurements of Refs. \cite{NANO1,PPTA,EPTA,IPTA} might 
not have anything to do with relic gravitons: the correlation signature of an isotropic gravitational wave background follows the so-called Hellings and Downs curve  which depends on the angle 
between a pair of Earth-pulsars baselines. As already mentioned this correlation has not been observed yet by admission 
of the various experimental collaborations \cite{NANO1,PPTA,EPTA,IPTA}. If we assume 
that the PTA are indeed related with relic gravitons the constraint of Eqs. (\ref{CONS2})--(\ref{NOT2})
is anyway more restrictive than the results of Eq. (\ref{PTAb1}). 
Specialising, for simplicity, to the case of scale-invariant flat spectrum we have that from Eq. (\ref{CONS2}):
\begin{equation}
h_{0}^2 \Omega_{gw}(\nu_{L}) < 10^{-8.61} \biggl(\frac{h_{0}}{0.65}\biggr)^2,
\label{RESTR1}
\end{equation}
where the limit has been referred to a fiducial value of $h_{0}$ that follows from the CMB data \cite{RT1,RT2,RT3}. 
By comparing Eqs. (\ref{PTAb1}) and (\ref{RESTR1}) it seems that the former is superficially 
more constraining than the latter: by choosing $q_{0} =1$ we would have that  $h_{0}^2\,\Omega_{gw}(\nu) < 10^{-8.86}$.
However $q_{0}$ is not $1$; on the contrary if we take the average of the four measurements 
presented so far (see Tab. \ref{TABLE2} in the case $\beta= -2/3$) we obtain and averaged value 
given by $\overline{q}_{0} = 2.467$ which implies 
\begin{equation} 
10^{-9.09} \biggl(\frac{\overline{q}_{0}}{2.467}\biggr)^2 \leq h_{0}^2 \, \Omega_{gw}(\nu) \leq 10^{-8.07} \biggl(\frac{\overline{q}_{0}}{2.467}\biggr)^2.
\label{RESTR2}
\end{equation}
Note that Eq. (\ref{RESTR1}) is always {\em more} constraining than Eq. (\ref{RESTR2}) 
even if we choose the smallest value of $q_{0}$ which is the one 
associated with the NANOgrav estimate \cite{NANO1}: if $q_{0} =1.92$ we get from Eq. (\ref{RESTR2}) that 
$h_{0}^2 \Omega_{gw}(\nu) \leq 10^{-8.29}$ which is always larger than the value of Eq. (\ref{RESTR1}). 
So far we simply considered the absolute values of the bounds but the frequency dependence 
of the theoretical spectra also matters: since $h_{0}^2 \, \Omega_{gw}(\nu,\tau_{0})$ generally increases 
with $\nu$, the limits of Eqs. (\ref{CONS2}) and (\ref{RESTR1}) are comparatively even more constraining 
that the ones of Eqs. (\ref{PTAb1}) and (\ref{RESTR2}) for the simple reason that $\nu_{L} = {\mathcal O}(60)$ Hz
while $\nu_{P} = {\mathcal O}(30)$ nHz. At high frequencies the limits of the audio band compete 
with the ones of nucleosynthesis (see the discussion hereunder). However if $h_{0}^2 \, \Omega_{gw}(\nu,\tau_{0})$
is nearly scale-invariant the most constraining bounds remain the ones associated with  Eqs. (\ref{CONS2})--(\ref{NOT2}) and (\ref{RESTR1}).

\subsection{Big-bang nucleosynthesis limits}
While the PTA measurements constrain the spectral energy density at intermediate 
frequencies, the bounds coming from big-bang nucleosynthesis \cite{bbn1,bbn2,bbn3} imply a constraint on the integral $h_{0}^2\,\Omega_{gw}(\nu,\tau_{0})$:
\begin{equation}
h_{0}^2  \int_{\nu_{bbn}}^{\nu_{max}}
  \Omega_{gw}(\nu,\tau_{0}) d\ln{\nu} = 5.61 \times 10^{-6} \Delta N_{\nu} 
  \biggl(\frac{h_{0}^2\,\Omega_{\gamma0}}{2.47 \times 10^{-5}}\biggr),
\label{CC2}
\end{equation}
where $\Omega_{\gamma0}$ is the (present) critical fraction of CMB photons. The limit  of Eq. (\ref{CC2}) sets an indirect constraint  on the extra-relativistic species possibly present at the time of nucleosynthesis. Since Eq. (\ref{CC2}) is relevant in the context of neutrino physics, the limit is often expressed for practical reasons  in terms of $\Delta N_{\nu}$ representing the contribution of supplementary neutrino species. The actual bounds on $\Delta N_{\nu}$ range from $\Delta N_{\nu} \leq 0.2$ 
to $\Delta N_{\nu} \leq 1$;  the integrated spectral density in Eq. (\ref{CC2}) is thus between $10^{-6}$ and $10^{-5}$. It is relevant to point out, as 
we shall see, that the upper limit of integration (labeled by $\nu_{max}$) depends on the specific 
post-inflationary evolutions\footnote{In the forthcoming discussion an important 
element is the determination of $\nu_{max}$ that depends on the duration of the post-inflationary 
evolution and on the corresponding expansion rates. For $\nu> \nu_{max}$ the 
spectra of relic gravitons are exponentially suppressed since these wavelengths 
never cross the Hubble radius and are not amplified.}. Conversely, the lower limit of integration in Eq. (\ref{CC2}) is given by  the frequency corresponding to the Hubble rate at the nucleosynthesis epoch: 
\begin{equation}
\nu_{bbn}= 2.252\times 10^{-11} \biggl(\frac{N_{eff}}{10.75}\biggr)^{1/4} \biggl(\frac{T_{bbn}}{\,\,\mathrm{MeV}}\biggr) 
\biggl(\frac{h_{0}^2\,\Omega_{R0}}{4.15 \times 10^{-5}}\biggr)^{1/4}\,\,\mathrm{Hz} \simeq 0.01\, \mathrm{nHz},
\label{CC3}
\end{equation}
where  $N_{eff}$ denotes the effective number of relativistic degrees of freedom entering the total energy density of the plasma and $T_{bbn}$ is the temperature of big-bang nucleosynthesis.  We finally remark that the bound of Eq. (\ref{CC2}) 
could be relaxed if the nucleosynthesis takes place in the presence 
of matter-antimatter domains \cite{bbn2}. This possibility 
will not be specifically considered hereunder and we shall instead enforce 
the bound of Eqs. (\ref{CC2})--(\ref{CC3}) in its conservative version.
As we shall see when the quasi-flat spectrum is normalized in the audio band 
the limit (\ref{CC2}) is always satisfied.

\renewcommand{\theequation}{4.\arabic{equation}}
\setcounter{equation}{0}
\section{Conditions for flat spectra at high-frequency}
\label{sec4}

\begin{figure}[!ht]
\centering
\includegraphics[height=7cm]{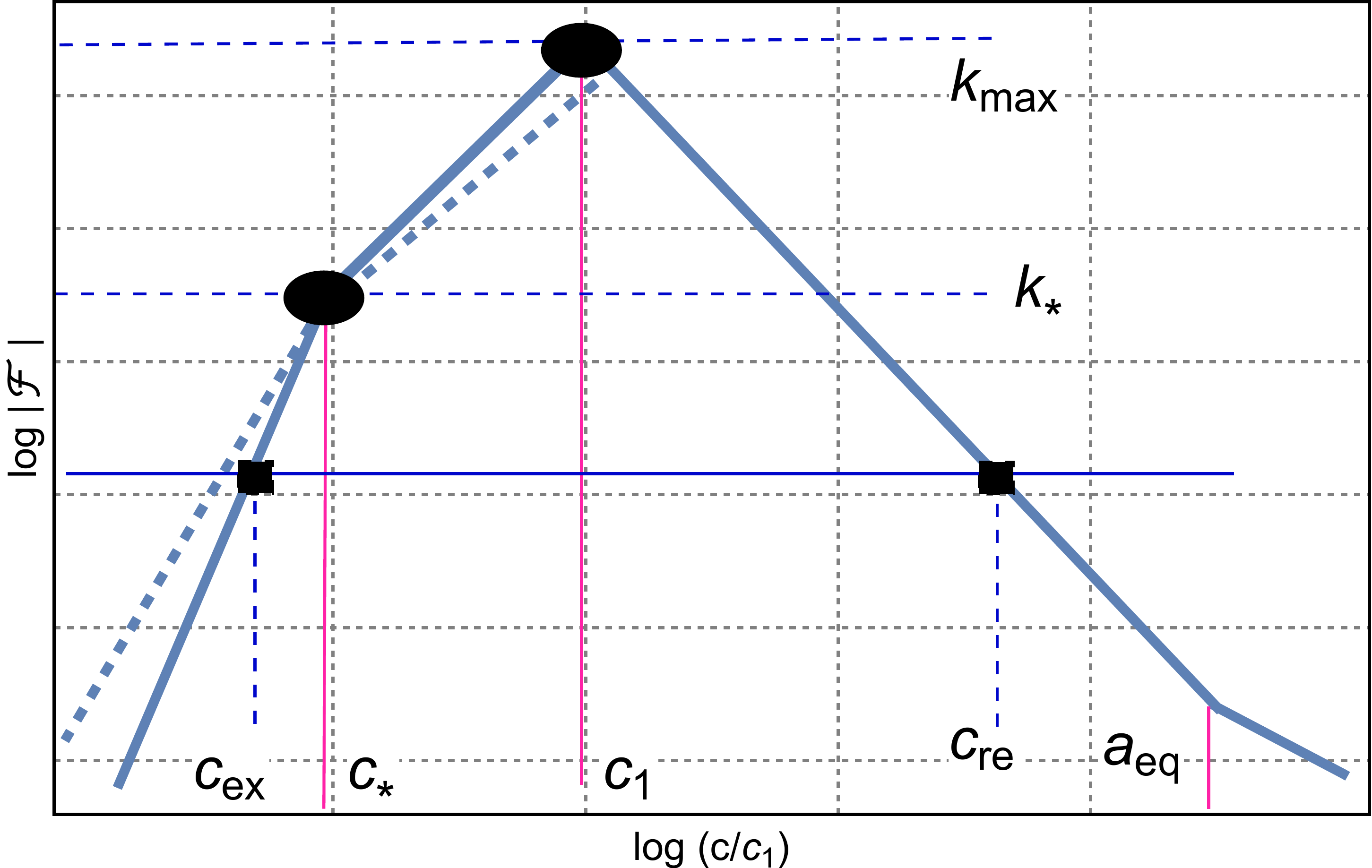}
\caption[a]{We schematically illustrate the evolution of ${\mathcal F}$  when the conventional 
inflationary phase is preceded by a further evolutionary stage. }
\label{FIG1}      
\end{figure}

In the conventional situation the spectral energy density for typical frequencies larger than the nHz is 
always smaller than $10^{-15}$. If all the sources of late-time 
suppression are taken into account we approximately have $h_{0}^2 \Omega_{gw}(\nu) = {\mathcal O}(10^{-16.5})$. This conclusion can be however evaded in, at least, two complementary 
situations that are simultaneously illustrated in Fig. \ref{FIG1} where the common logarithm of $|\,{\mathcal F}=\dot{b}/b\,|$ is reported.
The blobs appearing in the cartoon represent the transition regimes between the different stages and they are immaterial 
in the approach based on WKB approximation that is adopted in the present section. 
Figure \ref{FIG1} illustrates two of the three limits discussed in Eqs. (\ref{LIMIT1})--(\ref{LIMIT2})
and (\ref{LIMIT3}). In the case $n\to 1$ the $\eta$-time coincides with the conformal time coordinate 
and ${\mathcal F} = {\mathcal H} = a \, H$. Furthermore for $a > a_{*}$ the 
refractive index is not dynamical, and, in this limit, $c = a$. The most conservative case is the one associated 
with Eq. (\ref{LIMIT3}) where the refractive index evolves and we shall consider, in particular, 
the situation where $n(a)$ increases and then gets back to $1$ for $c > c_{\ast}$ (see Eq. (\ref{NEX}) and discussion therein).  

For $k < k_{\ast}$ the amplitude and the slope of $\Omega_{gw}(k,\tau)$  depend on the profile 
of ${\mathcal F}$ for $c< c_{\ast}$;  this means that for  $k_{eq} < k < k_{\ast}$, 
$\Omega_{gw}(k,\tau)$ increases provided the rate of variation of $n$ is sufficiently large in Hubble units. In the language 
of Eq. (\ref{calF3}) this implies $\alpha >0$ even if, as we shall see, 
$\alpha$ cannot exceed $1$ if the results of the wide band detectors are used to set 
the high-frequency normalization of the spectral energy density. 
According to Fig. \ref{FIG1} the spectral range $k_{\ast} < k < k_{max}$ is determined 
by the evolution of ${\mathcal F}$ for $c_{\ast} < c < c_{1}$ where ${\mathcal F} \propto a^{-1}$. 
This scaling actually corresponds to the conventional situation since, in this interval, ${\mathcal F} \propto {\mathcal H} \simeq a H$ so that  the Hubble rate is roughly constant.
In the minimal situation where the refractive index flattens out for $c > c_{\ast}$ the spectral 
energy density is quasi-flat (or slightly decreasing) simply because, in this range,
we get back to the conventional case where $\Omega_{gw}(k,\tau)$ is determined 
by the wavelengths crossing the Hubble radius during the inflationary stage and 
reentering when the background is dominated by radiation. Indeed for $c>c_{1}$ 
the evolution is completely standard and ${\mathcal F} = {\mathcal H} \propto a^{-1}$.
Finally for $c > a_{eq}$ we have that ${\mathcal F} = {\mathcal H} \propto a^{-1/2}$. 
In Fig. \ref{FIG1} we did not include the stage dominated by the dark energy 
simply because its contribution to the spectral energy density is immaterial for the 
considerations of this section. A more detailed account can be however found in section \ref{sec5}
where the late-time suppression of the spectral energy density will be more specifically investigated.
For the normalization in the high-frequency regime it is useful to have a fairly 
general expression of the spectral energy density that can be deduced rather 
simply within the WKB approximation; a key role is played, in this context, by the structure of the 
turning points $\eta_{ex}$ and $\eta_{re}$. As illustrated in Fig. \ref{FIG1}, a generic wavenumber 
$k< k_{max}$ crosses $|{\mathcal F}(\eta)|$ twice in $c_{ex}$ and $c_{re}$ and these two values 
correspond to the moments where a given wavelength exists and reenters the effective Hubble 
radius $| {\mathcal F}(\eta) |^{-1}$.  From a technical viewpoint  $\eta_{ex}$ and $\eta_{re}$ are defined as the turning point at which the solution to Eq. (\ref{NM16}) changes its analytic form. In particular the exit corresponds to: 
\begin{equation}
k^2 = \biggl|\frac{\ddot{c}_{ex}}{c_{ex}}\biggr|, \qquad \ddot{c}_{ex} \neq 0 \qquad \Rightarrow\qquad k = 
\sqrt{\biggl|\frac{\ddot{c}_{ex}}{c_{ex}}\biggr|}.
\label{EQAP2}
\end{equation}
Equation (\ref{EQAP2}) actually defines a regular turning point and it tells that, at $\eta_{ex}$, $k \eta_{ex} \simeq 1$. It can happen, however, that the turning point is singular 
and this happens, in particular, when $\ddot{c} \to 0$ in the vicinity of the turning point. In the problem at hand 
we have that $\ddot{c}_{ex} \neq 0$ however, if the reentry takes place during a radiation-dominated 
stage of expansion, we typically have $\ddot{c}_{re} \to 0$ since, in a radiation stage, $\eta=\tau$ and 
$a^{\prime\prime}=0$. In this case, as we shall see in a moment, we must recall that the condition 
$k \eta_{re} \simeq 1$ is not verified and it must be replaced by $ k \eta_{re} \ll  1$. In what 
follows we shall first discuss the expression of the spectral energy density for the different 
ranges of wavelengths and then analyze the way the high-frequency normalization 
must be implemented.

\subsection{Wavelengths larger than the effective horizon}
To deduce the general evolution of the mode function for large wavelengths it is practical 
to transform the relevant differential equations in a set of integral equations.
The evolution of the mode function of Eq. (\ref{NM16}) is equivalent to\footnote{The same conclusion also holds in the case of Eq. (\ref{NM14}) that is valid in the limit (\ref{LIMIT1}). The conditions (\ref{LIMIT3}) and (\ref{NM16}) are actually more general than (\ref{LIMIT1}).} an integral equation whose initial conditions are assigned at the reference time $\eta_{ex}$:
\begin{equation}
f_{k}(\eta) = \frac{c(\eta)}{c_{ex}} \biggl\{ f_{k}(\eta_{ex}) 
+ g_{k}(\eta_{ex}) \int_{\eta_{ex}}^{\eta} \frac{c_{ex}^2}{c^2(\eta_{1})} d\eta_{1}
- k^2 \, c_{ex}\, \int_{\eta_{ex}}^{\eta} \frac{d\eta_{1}}{c^2(\eta_{1})} \int_{\eta_{ex}}^{\eta_{1}} \,c(\eta_{2}) \,f_{k}(\eta_{2}) d\eta_{2} \biggr\},
\label{EQAP1}
\end{equation}
where, by definition,  $c_{ex} = c(\eta_{ex})$. As argued in Eq. (\ref{EQAP2}) 
the first turning point is always regular so that $k \eta_{ex} \simeq 1$.
Neglecting then the terms ${\mathcal O}(k^2 \eta^2)$ the lowest order solution of Eq. (\ref{EQAP1}) is:
\begin{equation}
f_{k}(\eta) = \frac{c(\eta)}{c_{ex}} \biggl\{ f_{k}(\eta_{ex}) 
+ g_{k}(\eta_{ex}) \int_{\eta_{ex}}^{\eta} \frac{c_{ex}^2}{c^2(\eta_{1})} d\eta_{1}\biggr\} + {\mathcal O}(k^2 \eta^2).
\label{MF3}
\end{equation}
Equation (\ref{MF3}) determine the approximate form of the power spectrum for wavelengths larger than the Hubble radius. Since  the second term appearing inside the squared bracket at the right hand side of Eq. (\ref{MF3}) is subleading for typical wavelengths larger than the effective horizon, the explicit expression 
of the tensor power spectrum follows from Eq. (\ref{calF17}) by recalling that, for $\eta < -\eta_{\ast}$, 
\begin{equation}
c(\eta) = c_{\ast} (-\eta/\eta_{\ast})^{- \zeta}, \qquad\qquad \zeta = (2- \alpha)/2( 1 + \alpha - \epsilon), \qquad\qquad c_{\ast} = a_{\ast}/\sqrt{n_{\ast}}.
\end{equation}
  The expression of $c(\eta)$ follows, in its turn, from the definition of the $\eta$-time 
and from Eq. (\ref{NEX}) evaluated in a conventional inflationary background 
with slow-roll rate given by $\epsilon$. In the limit $|\, k\, \eta\,| \ll 1$ the tensor power spectrum is in fact constant and it is approximately given by:
\begin{equation}
P_{T}(k,\eta) = \frac{2 \, \ell_{P}^2 }{ \pi^2 \, c_{*}^2 \,\eta_{*}^2} |A(\alpha,\epsilon)|^2  \, (- k \eta_{*})^{n_{T}(\alpha,\epsilon)}, \qquad\qquad k < k_{\ast},
\label{PS1} 
\end{equation}
where $A(\alpha,\epsilon)$ and $n_{T}(\alpha,\epsilon)$ are defined, respectively, by:
\begin{equation}
n_{T} = \frac{3 \alpha - 2 \epsilon}{1 -\epsilon + \alpha}, \qquad |A(\nu)| = \frac{\Gamma(\nu)}{\sqrt{\pi}} \, 2^{\nu - 1/2}, \qquad \nu = \ \frac{3  - \epsilon}{2 ( 1 + \alpha - \epsilon)}.
\label{PS4}
\end{equation}
Within the approximation scheme leading to Eq. (\ref{MF3}) we have that in Eq. (\ref{PS1}) $\bigl| A \bigr|= \sqrt{2k} \,\,\bigl| f_{k}(\eta_{ex}) \bigr| =1$. In a more general perspective the amplitude $\bigl| A \bigr|$ appearing in Eq. (\ref{PS1}) parametrizes, up to an irrelevant phase, the mismatch between the exact and the approximate 
solutions at $\eta_{ex}$: for $k^2 \ll |\ddot{c}/c|$ the correctly normalized solutions of Eq. (\ref{NM14}) are $f_{k}(\eta) = e^{\pm \,i k\eta}/\sqrt{2 k}$. However as soon as $\eta_{ex}$ is approached the amplitude gets slightly modified and the exact solution of Eq. (\ref{NM14}) is given in terms of Hankel functions \cite{abr1,abr2}. The power spectrum of  Eq. (\ref{PS1}) can be further simplified by observing that: 
\begin{equation}
\frac{1}{\eta_{\ast}^2} = \frac{n_{\ast}^2}{\tau_{*}^2} \biggl|1 - \frac{\alpha}{1-\epsilon}\biggr|^2 \qquad \Rightarrow\qquad 
\frac{1}{\eta_{\ast}^2} = a_{\ast}^2 \, n_{\ast}^2 \, H_{\ast}^2 \biggl|1 - \frac{\alpha}{1-\epsilon}\biggr|^2.
\label{PS4a}
\end{equation}
If we now use Eqs. (\ref{PS4})--(\ref{PS4a}) into Eqs. (\ref{PS1}) we obtain the following form of the power 
spectrum for $k < a_{\ast} \, H_{\ast}$
\begin{equation}
P_{T}(k,\eta_{*}) = \biggl(\frac{H_{*}}{M_{P}}\biggr)^2\,\, \frac{2^{6 - n_{T}}}{\pi^2} \,\Gamma^2\biggl(\frac{3 - n_{T}}{2}\biggr) \, n_{*}^{ 3 - n_{T}}\, 
\biggl| 1 + \frac{\alpha}{1 - \epsilon}\biggr|^{2 -n_{T}} \,\,\biggl( \frac{k}{a_{*} H_{*}}\biggr)^{n_{T}}.
\label{PS5}
\end{equation}
For the modes $k > a_{\ast} \, H_{\ast}$ the spectrum (\ref{PS5}) is modified since now the evolution 
of the mode functions is given by:
\begin{equation}
f_{k}^{\prime\prime} + \biggl[\omega^2 - \frac{a^{\prime\prime}}{a} \biggl] f_{k} =0, \qquad \omega^2 = k^2/n_{\ast}^2.
\label{PS5a}
\end{equation}
Recalling that $a^{\prime\prime}/a = a^2 H^2 (2 - \epsilon)$ and that $a\, H = - 1/[ (1 -\epsilon)\tau]$ 
Eq. (\ref{PS5a}) becomes 
\begin{equation}
f_{k}^{\prime\prime} + \biggl[\omega^2 -\frac{3 - \epsilon}{2( 1-\epsilon)^2}\biggl] f_{k} =0, 
\label{PS5b}
\end{equation}
If the refractive phase terminates before the end of inflation 
the power spectrum inherits a further branch for $a_{*} H_{*} < k \leq a_{1} H_{1}$:
\begin{equation}
 P_{T}(k,\tau_{1}) = \biggl(\frac{H_{1}}{M_{P}}\biggr)^2\,\,n_{*}^{3 -m_{T}}\, \frac{2^{6 - m_{T}}}{\pi^2} \,\Gamma^2\biggl(\frac{3 - m_{T}}{2}\biggr)  \,\,\biggl( \frac{k}{a_{1} H_{1}}\biggr)^{m_{T}},\qquad m_{T} = - 2\epsilon/(1 - \epsilon).
\label{PS5c}
\end{equation}

\subsection{Wavelengths shorter than the effective horizon}
All the cosmic gravitons measured at the present time are inside the Hubble radius 
and provided the reentry occurs when $\ddot{c}_{re} \neq 0$ we have that, approximately, 
 $k \eta_{re} = {\mathcal O}(1)$. Conversely 
if $\ddot{b}_{re} \to 0$ in the vicinity of the turning point, then $k \eta_{re} \ll 1$. For $\eta \geq \eta_{re}$ the solution of Eq. (\ref{NM14}) is
\begin{equation}
f_{k}(\eta) = {\mathcal C}_{+}(k,\eta_{ex},\eta_{re}) \,\,\overline{f}_{re}(\eta) + {\mathcal C}_{-}(k,\eta_{ex},\eta_{re}) \,\,\overline{f}_{re}^{*}(\eta),
\label{MF5a}
\end{equation}
where $\overline{f}_{re}(\eta)$ are the mode functions inside the effective horizon (i.e. $e^{- i k\eta}/\sqrt{2 \, k}$ in the crudest approximation). The coefficients ${\mathcal C}_{\pm}(k,\eta_{ex},\eta_{re})$ are 
\begin{eqnarray}
{\mathcal C}_{\pm}(k,\eta_{ex},\eta_{re}) &=& \frac{e^{- i\, k ( \eta_{ex} \mp \eta_{re})}}{2\, i\, k} \biggl[ \pm \frac{c_{ex}}{c_{re}}({\mathcal F}_{ex} + i \, k) \mp \frac{c_{re}}{c_{ex}} ({\mathcal F}_{re} \mp i k) 
\nonumber\\
&\pm& c_{re} \, c_{ex} ({\mathcal F}_{ex} + i k) ({\mathcal F}_{re}\mp i\, k) {\mathcal J}(\eta_{ex}, \, \eta_{re})\biggr],
\nonumber\\
{\mathcal J}(\eta_{ex}, \, \eta_{re}) &=& \int_{\eta_{ex}}^{\eta_{re}} \frac{d\eta}{c^2(\eta)}.
\label{MF5b}
\end{eqnarray}
If the reentry takes place, as we are considering here, when the refractive index is not 
dynamical, Eq. (\ref{MF5b}) can be simplified even further since $b(\eta) \to a(\tau)$ and 
${\mathcal F} \to {\mathcal H}$:
\begin{eqnarray}
{\mathcal C}_{\pm}(k,\eta_{ex},\tau_{re}) &=& \frac{e^{- i\, k ( \eta_{ex} \mp \tau_{re})}}{2\, i\, k} \biggl[ \pm \frac{c_{ex}}{a_{re}}({\mathcal F}_{ex} + i \, k) \mp \frac{a_{re}}{c_{ex}} ({\mathcal H}_{re} \mp i k) 
\nonumber\\
&\pm& a_{re} \, c_{ex} ({\mathcal F}_{ex} + i k) ({\mathcal H}_{re}\mp i\, k) {\mathcal J}(\eta_{ex}, \, \eta_{re})\biggr],
\label{MF5ca}\\
{\mathcal J}(\eta_{ex}, \, \tau_{re}) &=& \int_{\eta_{ex}}^{-\tau_{1}} \frac{d\eta}{c^2(\eta)} + \int_{-\tau_{1}}^{\tau_{re}} \frac{d\eta}{a^2(\tau)},
\label{MF5cb}
\end{eqnarray}
where $-\tau_{1}$ marks, as before, the end of the inflationary stage.
Because $c(\eta)$ always increases, 
in Eqs. (\ref{MF5ca})--(\ref{MF5cb})the terms proportional to $|c_{ex}/c_{re}|$ can be neglected in comparison with $|b_{re}/b_{ex}|$. Since ${\mathcal C}_{\pm}(k)$ are both complex but subjected to the condition 
$\bigl|{\mathcal C}_{+}(k,\eta_{ex},\eta_{re})\bigr|^2 - \bigl|{\mathcal C}_{-}(k,\eta_{ex},\eta_{re})\bigr|^2 =1$ it is sufficient to estimate the approximate form of $|{\mathcal C}_{-}(k,\eta_{ex},\eta_{re})|^2$:
\begin{equation}
|{\mathcal C}_{-}(k,\eta_{ex},\eta_{re})|^2 \simeq \frac{1}{4} \biggl(\frac{c_{re}}{c_{ex}}\biggr)^2 \biggl(1 + \frac{{\mathcal F}_{re}^2}{k^2} \biggr) \biggl[ 1 - 2 {\mathcal F}_{ex} c_{ex}^2 {\mathcal J}(\eta_{ex}, \eta_{re})
+ c_{ex}^4 ( {\mathcal F}_{ex}^2 + k^2) {\mathcal J}^2(\eta_{ex}, \eta_{re})\biggr].
\label{MF5d}
\end{equation}
Equation (\ref{MF5d}) allows for a swift determination of the power spectrum and of the 
spectral energy distribution in the limit $k\tau \gg 1$, i.e. when the relevant wavelengths are all inside the 
Hubble radius:
\begin{eqnarray}
P_{T}(k,\tau) &=& \frac{4 k^2}{\pi^2 \overline{M}_{P}^2\, a^2} \bigl| {\mathcal C}_{-}(k,\eta_{ex},\eta_{re})\bigr|^2 \biggl[ 1 + {\mathcal O} \biggl(\frac{1}{k^2 \tau^2}\biggr) \biggr], 
\label{MF5e}\\
\Omega_{gw}(k,\tau) &=& \frac{k^4}{3 H^2 \overline{M}_{P}^2 \pi^2 a^4} \bigl| {\mathcal C}_{-}(k,\eta_{ex},\eta_{re})\bigr|^2 \biggl[ 1 + {\mathcal O} \biggl(\frac{1}{k^2 \tau^2}\biggr) \biggr], 
\label{MF5f}
\end{eqnarray} 
From the ratio between Eqs. (\ref{MF5e})--(\ref{MF5f})  the standard relation 
between the power spectrum and the spectral energy density is recovered
\begin{equation}
\Omega_{gw}(k,\tau) = \frac{k^2}{12 a^2 H^2} P_{T}(k,\tau) \biggl[ 1 + {\mathcal O} \biggl(\frac{1}{k^2 \tau^2}\biggr) \biggr],
\label{MF5g}
\end{equation}
and it is generally valid when the relevant wavelengths are shorter than the Hubble radius at a given epoch.
Inside the Hubble radius we can evaluate indifferently either the power spectrum or the spectral energy distribution. It is finally useful to estimate more explicitly $k_{\ast}$ and $k_{max}$. Since $k_{\ast} = 1/\eta_{\ast}$ we also have that 
\begin{equation}
k_{\ast} = \biggl| 1 + \frac{\alpha}{1-\epsilon}\biggr|\,\, e^{\alpha\, N_{\ast}} e^{- \Delta N} k_{max}, \qquad\qquad 
\Delta N= N_{t} - N_{\ast}.
\label{MF5h}
\end{equation}
As usual in Eq. (\ref{MF5h}) $N_{\ast} = \ln{(a_{\ast}/a_{i})}$ denotes the number 
of $e$-folds during the refractive stage  $N_{t} = \ln{(a_{1}/a_{i})}$ is the total number of $e$-folds. 
Recalling that $\nu_{max} = k_{max}/(2\pi)$ we then have, in explicit terms:
\begin{equation}
\nu_{max}= 269.33 \,\biggl(\frac{\epsilon}{0.003}\biggr)^{1/4} \,\biggl(\frac{{\mathcal A}_{{\mathcal R}}}{2.41\times 10^{-9}}\biggr)^{1/4} \, \biggl(\frac{h_{0}^2 \, \Omega_{R0}}{4.15 \times 10^{-5}}\biggr)^{1/4} \,\,\,\mathrm{MHz},
\label{MF5i}
\end{equation}
where ${\mathcal A}_{{\mathcal R}}$ is the amplitude of the scalar power spectrum at the pivot scale $k_{p} = 0.002\,\, 
\mathrm{Mpc}^{-1}$ and $\Omega_{R0}$ is the total fraction of relativistic species at the present time 
in the concordance paradigm. Thanks to Eq. (\ref{MF5h}) we also have, by definition, that 
\begin{equation}
\nu_{\ast} = \biggl( 1 + \frac{\alpha}{1-\epsilon}\biggr)\,\, e^{\alpha\, N_{\ast}} e^{- \Delta N} \,\nu_{max}.
\label{MF5l}
\end{equation}
We note that in Eq. (\ref{MF5i}) we introduced the slow-roll parameter 
$\epsilon$ and not $r_{T}$ since we did not assume the consistency 
relations so that $r_{T}$ might be smaller (or even much smaller) 
than $0.06$.
\subsection{Flat spectra at high frequencies}
The spectral energy density for typical wavenumbers $ a_{*} \, H_{*} < k < a_{1}\, H_{1}$ 
is quasi-flat and to show this point in general terms it is useful to go back to Eqs. (\ref{MF5d}) and (\ref{MF5f}):
\begin{equation}
\Omega_{gw}(k,\tau) = \frac{k^4}{12 \,\pi^2\, H^2\, \overline{M}_{P}^2\, a^{4}} \biggl|\frac{c_{re}(k)}{c_{ex}(k)}\biggr|^2 \biggl(
1 + \frac{1}{k^2 \tau_{re}^2} \biggr).
\label{OGW1}
\end{equation}
As already mentioned twice, in Eq. (\ref{OGW1}) we have two complementary possibilities depending on the nature of the turning point\footnote{
If the reentry takes place during the radiation phase we have that, in the vicinity of $\tau_{re}$ 
$a^{\prime\prime} \to 0$ so that $k \tau_{re}\ll 1$ in Eq. (\ref{OGW1}). Conversely the modes
exit during the inflationary phase when $\omega^2 \tau_{ex} \simeq 1$ (i.e. $k^2 \, \tau_{ex}^2 \simeq n_{*}^2$).}.  When the wavelength renters the Hubble radius during radiation the correct limit is $k \tau_{re} \ll 1$ 
and the spectral energy density at high-frequency can then be estimated as:
\begin{equation}
\Omega_{gw}(k,\tau_{0}) = \frac{k^2 \, n_{\ast}}{12 \, \pi^2 \, a_{ex}^2 \, \overline{M}_{P}^2} \biggl(\frac{H_{re}^2 \, a_{re}^{4}}{H_{0}^2 a_{0}^4}\biggr).
\label{OGW2}
\end{equation}
For the slice of wavenumbers $ a_{*} \, H_{*} < k < a_{1}\, H_{1}$ the exit takes place after the refractive 
phase where $a_{ex} = n_{\ast}^{- 1/(1 -\epsilon)} \, |k\,\tau_{1}|^{1/(1 - \epsilon)}$. From Eq. (\ref{OGW2}) 
we then obtain:
\begin{eqnarray}
h_{0}^2 \Omega_{gw}(\nu) &=& \biggl(\frac{H_{1}}{M_{P}}\biggr)^2 \overline{\Omega}_{\ast} \biggl(\frac{\nu}{\nu_{\ast}}\biggr)^{m_{T}}, \qquad\qquad \nu_{\ast} < \nu < \nu_{max},
\label{OGW3}\\
\overline{\Omega}_{\ast} &=& \frac{4 \, h_{0}^2 \Omega_{R0}}{3 \pi} \, n_{\ast}^{3} \, \biggl( 1 + \frac{\alpha}{1-\epsilon}\biggr)^{m_{T}}\,\, e^{- m_{T} \, \Delta N},
\label{OGW4}
\end{eqnarray}
where, as usual, $\Delta N= (N_{t} - N_{\ast})$. The same reasoning for lower frequencies leads to: 
\begin{eqnarray}
h_{0}^2 \Omega_{gw}(\nu) &=& \biggl(\frac{H_{1}}{M_{P}}\biggr)^2 \overline{\Omega}_{\ast} \,\,\biggl(\frac{\nu}{\nu_{\ast}}\biggr)^{n_{T}}, \qquad\qquad\qquad \nu_{eq} < \nu < \nu_{\ast},
\label{OGW5}\\
h_{0}^2 \Omega_{gw}(\nu) &=& \biggl(\frac{H_{1}}{M_{P}}\biggr)^2 \overline{\Omega}_{\ast}\,\, \biggl(\frac{\nu}{\nu_{\ast}}\biggr)^{n_{T}}\, \,\biggl(\frac{\nu}{\nu_{eq}}\biggr)^{-2}, \qquad \nu < \nu_{eq}.
\label{OGW6}
\end{eqnarray}
Equations (\ref{OGW3})--(\ref{OGW4}) and (\ref{OGW5})--(\ref{OGW6}) give $\Omega_{gw}(k,\tau)$ 
in the three spectral regions defined in Fig. \ref{FIG1}. In summary the lowest frequency region involves the modes 
exiting the Hubble radius during the refractive phase and reentering after equality (i.e. $k < a_{eq} H_{eq}$). 
The intermediate region concerns the modes exiting the effective horizon during the refractive phase and reentering 
during radiation (i.e. $a_{eq} H_{eq}< k < a_{*} H_{*}$). The highest frequency domain encompasses the modes that exit the Hubble radius after the end of the refractive phase and reenter during radiation 
(i.e. $a_{*} H_{*} < k < a_{1} H_{1}$). As we shall see the relevant physical regime is the one 
where the rate of variation of $n$ is larger than $\epsilon$;  in this limit we can expand the 
spectral index for $\epsilon \ll 1$:
\begin{equation}
n_{T} = \frac{3 \alpha - 2\epsilon}{(1 + \alpha - \epsilon)}=
\frac{3 \alpha }{1 + \alpha} + \frac{[ -2 + \alpha ( 1 - 2\gamma)]\epsilon}{(1 + \alpha)^2} + {\mathcal O}(\epsilon^2),
\label{sq8a}
\end{equation}
where the second equality follows in the limit $\epsilon \ll 1$. The spectra obtained so far in the case of a 
dynamical refractive index are fully compatible with a conventional inflationary stage 
and this is ultimately the reason for the flatness of $\Omega_{gw}(k,\tau)$ at high-frequencies. 
It is however interesting to mention that the  parametrization of  Eqs. (\ref{OGW3})--(\ref{OGW4}) and (\ref{OGW5})--(\ref{OGW6}) holds also, with the appropriate differences, when the spectrum at intermediate frequencies is not dictated by the evolution of the refractive index.  Two cases are particularly important in the light of the current data: the bounces of the scale factor and the curvature bounces. For a bounce of the scale factor the scale factor first contacts in an accelerated manner\footnote{This means that $\dot{a} <0$ and $\ddot{a} <0$ where, only in this paragraph, the overdot denotes a derivation 
with respect to the cosmic time coordinate $t$.} and then undergoes a stage of decelerated 
expansion (i.e. $\dot{a} >0$, $\ddot{a}<0$). In the case of an accelerated 
contraction the scale factor can be parametrized as a power-law
$a(t) \simeq (-t/t_{1})^{\delta}$ with $0< \delta < 1$.  When $\delta < 0$ we have instead an accelerated expansion with growing curvature (i.e. $\dot{H}>0$). The spectral energy density of the relic gravitons produced in this kind of scenarios can be estimated as in the previous case with the relevant 
similarity that the intermediate spectral index is also blue. In the case 
of accelerated contraction we have 
\begin{equation}
n_{T} = 3 - \biggl| \frac{2 \delta}{\delta -1} - 1\biggr|, \qquad\qquad 0< \delta <1.
\label{BOUNCE1}
\end{equation}
In the case of accelerated expansion with $\delta < 0$ we have instead 
\begin{equation}
n_{T} = \frac{2}{|\delta| +1}, \qquad\qquad \delta < 0.
\label{BOUNCE2}
\end{equation}
For $\nu> \nu_{*}$ the slope s controlled by $m_{T}$ even if, in this 
case, the quasi-flat slope is not related to a slow-roll dynamics as in the conventional inflationary case. We can therefore conclude that 
$\overline{\Omega}_{\ast}$ is generally given by the product of two separate 
contributions 
\begin{equation}
\overline{\Omega}_{\ast} = \overline{\Omega}_{late}(\Omega_{R0}, \Omega_{\Lambda}, N_{eff}, N_{\nu}) \, \, \overline{\Omega}_{early}
\label{OOO}
\end{equation}
where $\overline{\Omega}_{late}$ denotes the late-time 
contribution that only depends on the parameters of the concordance 
scenario; in Eq. (\ref{OOO}) $\overline{\Omega}_{early}$ is instead the early contribution that is generally model-dependent.
In particular we have that, in the present case, $\overline{\Omega}_{\ast} = \overline{\Omega}_{\ast}(\alpha, \epsilon, N_{\ast}, N_{t})$.

\subsection{High-frequency normalization of the spectral energy density}
The high-frequency normalization can be studied accurately by considering 
all the late-time sources of suppression but this analysis is postponed to the 
following section since, in what follows,  the attention is focused on the general 
logic that can be more directly appreciated from the analytic results deduced above. The first 
observation is that in the conventional situation $(H_{1}/M_{P})$ appearing in Eq. 
(\ref{OGW3})  is fixed from the amplitude of the scalar power spectrum ${\mathcal A}_{{\mathcal R}}$ and from the slow-roll parameter:
\begin{equation}
\biggl(\frac{H_{1}}{M_{P}}\biggr) = 4.7\times 10^{-6} \biggl(\frac{{\mathcal A}_{{\mathcal R}}}{2.41 \times 10^{-9}}\biggr)^{1/2}\,\,\biggl(\frac{\epsilon}{0.003}\biggr)^{1/2}.
\label{HFnorm1}
\end{equation}
When the normalization is set at low-frequencies $\epsilon$ is related to the 
$r_{T}$ whose upper limits fix the spectral energy density in the aHz range.
In the present case, however, the situation is different and there are, in purely abstract terms, two complementary possibilities:
\begin{itemize}
\item{} the first logical possibility is to fix the normalization by 
requiring that the spectral energy density  matches the value 
measured by the PTA at the typical frequency $\nu=\nu_{P} = {\mathcal O}(30) \, \mathrm{nHz}$;
\item{} the second possibility is instead to normalize the potential 
signal in the audio band by enforcing the LVK bound at the 
frequency $\nu= \nu_{L} = {\mathcal O}(60) \, \mathrm{Hz}$.
\end{itemize}
Between these two possibilities the latter is more plausible 
than the former  for the simple reason that, generally 
speaking, $ \nu_{*} < \nu_{L}$ so that the quasi-flat branch 
of the spectrum falls in the audio band.
\begin{figure}[!ht]
\centering
\includegraphics[height=8.5cm]{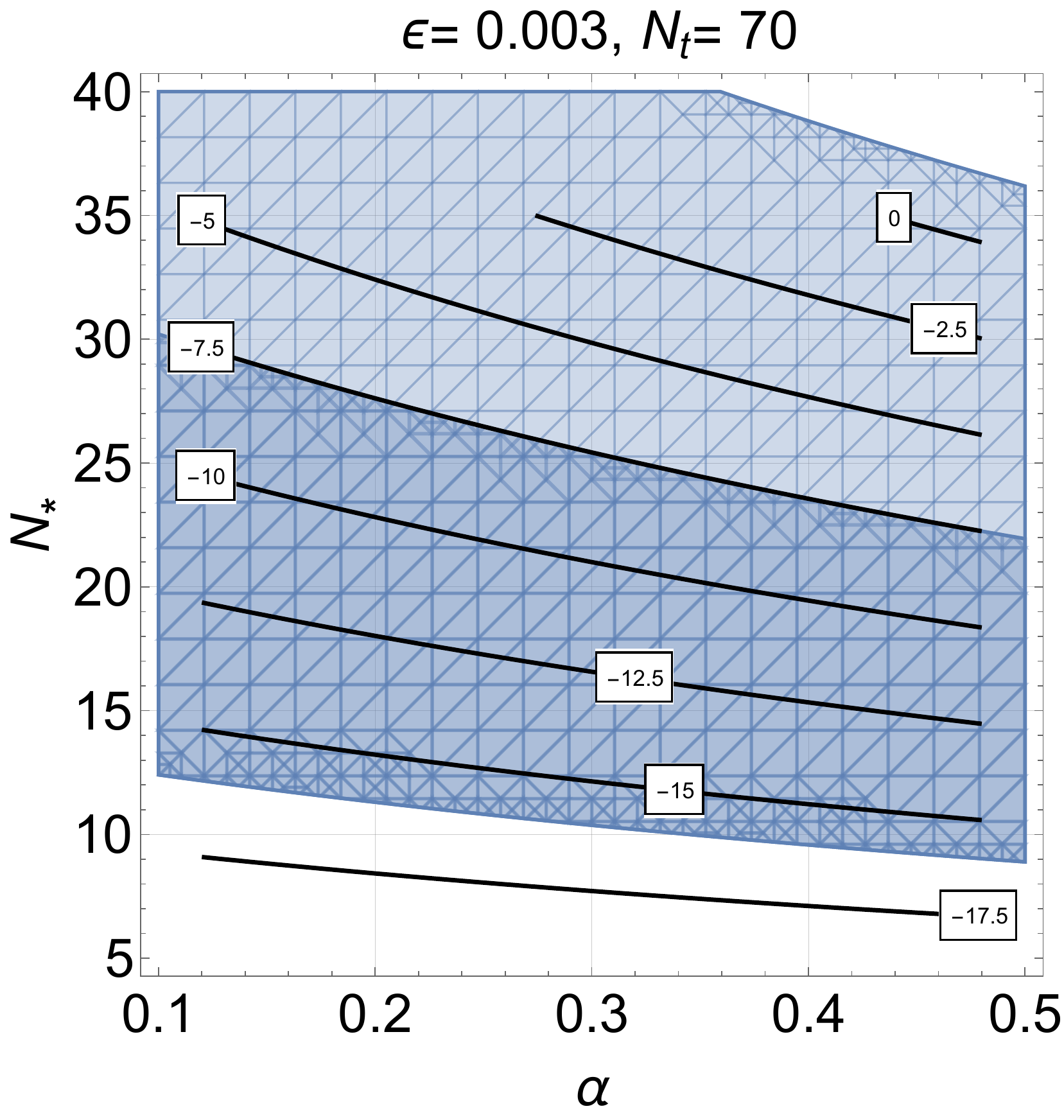}
\includegraphics[height=8.5cm]{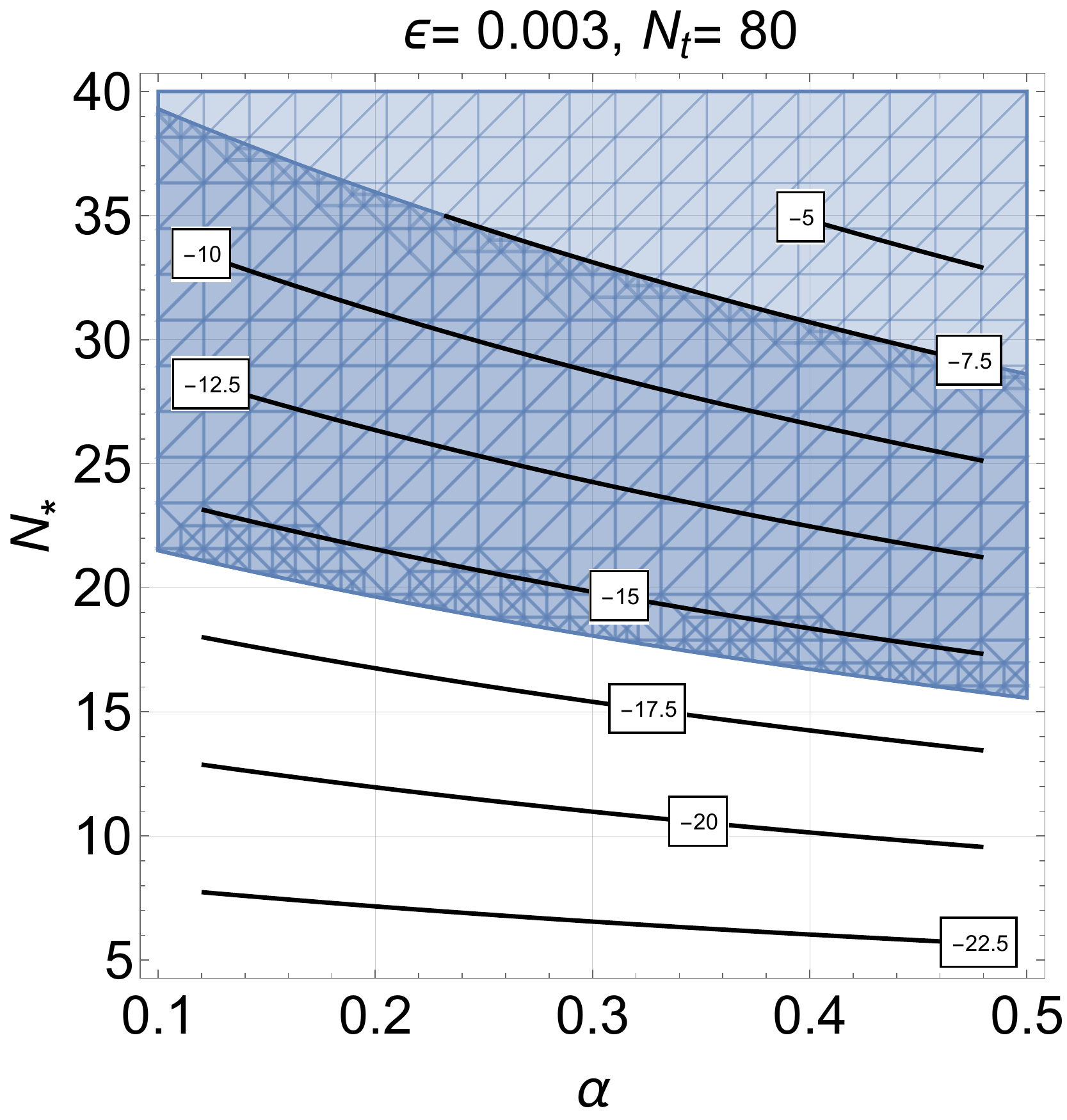}
\caption[a]{The shaded area illustrates the region of the parameter 
space where, according to Eq. (\ref{MF5l}), $100\, \mathrm{aHz}< \nu_{\ast} < \nu_{L}$. The
two plots differ because of the total number of $e$-folds but are otherwise 
qualitatively similar. The darker region in the central part of both plots corresponds 
to $100\, \mathrm{aHz} \nu_{\ast} < \nu_{P}$. The various curves are the 
contours where the values of $\nu_{\ast}$ remain the same and in the labels 
we report the common logarithm of $\nu_{\ast}$ expressed in Hz.}
\label{FIG2}      
\end{figure}
On the other hand it is plausible to have that $\nu_{\ast}$ is either 
larger or smaller than $\nu_{P}$.
This point is illustrated in Fig. \ref{FIG2} where the darker 
area defines the region where $100 \,\, \mathrm{aHz} <\nu_{\ast} < \nu_{P}$.
The various labels appearing on the curves define the common
logarithm of $\nu_{\ast}$ expressed in Hz. We have that, approximately, 
$\nu_{P} = {\mathcal O}(10^{-7.5})\, \mathrm{Hz}$.
Figure \ref{FIG2} explains why the high-frequency 
normalization has ben fixed by requiring 
\begin{equation}
\biggl(\frac{H_{1}}{M_{P}}\biggr)^2 \overline{\Omega}_{\ast} \biggl(\frac{\nu_{L}}{\nu_{\ast}}\biggr)^{m_{T}} = h_{0}^2 \overline{\Omega}_{gw}(\nu_{L}), \qquad\qquad \nu > \nu_{\ast}.
\label{HFnorm2}
\end{equation}
According to Eqs. (\ref{CONS2})--(\ref{NOT2}) the absolute upper limit 
for the quasi-flat spectral energy density should be ${\mathcal O}(5.8) \times 10^{-9}$
and, for the sake of simplicity, we are going to require, in a conservative perspective\footnote{This estimate ignores, by 
explicit choice,  the sources of damping that have been instead taken into account 
in section \ref{sec5}. It must be understood as an order of magnitude evaluation. In Fig. \ref{FIG4} the 
various later-time damping sources have been included and the obtained results are compared with the analytic 
estimates. In spite of the (obvious) slight numerical disagreement it turns out that the 
analytic results are quite useful to illustrate the general logic.}, 
that $\overline{\Omega}_{gw}(\nu_{L}) =5.8 \times 10^{-9}$. 
To illustrate the procedure (and to avoid the possible complications that are addressed 
in the next section) we assume, in the notations of Eq. (\ref{OOO}), that 
\begin{eqnarray}
&& \overline{\Omega}_{late}(\Omega_{R0}) = \frac{4 }{3\pi} \Omega_{R0},
\label{HFnorm3a}\\
&& \overline{\Omega}_{early}(\alpha, \epsilon, N_{\ast}, N_{t}) = 
\, e^{3 \alpha N_{\ast}} \, \biggl( 1 + \frac{\alpha}{1-\epsilon}\biggr)^{m_{T}}\,\, e^{- m_{T} \, \Delta N}.
\label{HFnorm3b}
\end{eqnarray}
From Eqs. (\ref{HFnorm2}) and (\ref{HFnorm3a})--(\ref{HFnorm3b}) the following relation can be deduced
\begin{equation}
e^{[3 - m_{T}(\epsilon)]\alpha\, N_{\ast}} \,\,e^{m_T(\epsilon) \Delta \, N} = \frac{3}{4} \biggl[ \frac{\Omega_{gw}(\nu_{L})}{\Omega_{R0} \, \epsilon \, {\mathcal A}_{{\mathcal R}}}\biggr]
\, \biggl(\frac{\nu_{max}}{\nu_{L}}\biggr)^{m_{T}(\epsilon)}.
\label{HFnorm4}
\end{equation}
As we mentioned in Eq. (\ref{CONS2}) the value of $\nu_{L}$ ranges approximately 
between $20$ and  $76$ Hz. This range is related to the frequency region which 
is more sensitive to the backgrounds of relic gravitons; for the sake of concreteness 
we therefore posit $\nu_{L} = 60$ Hz. To pass from Eq. (\ref{HFnorm4}) 
we first employed Eq. (\ref{HFnorm2}) and also noted that $\nu_{L}/\nu_{\ast} = (\nu_{L}/\nu_{max}) 
(\nu_{max}/\nu_{\ast})$ since the ratio $(\nu_{max}/\nu_{\ast})$ can be directly estimated 
from Eq. (\ref{MF5l}). In practice all the terms at the left of Eq. (\ref{HFnorm4}) contain the parameters of the model while the quantities ate the right-hand side are either directly measured 
or can be determined as late-time parameters of the concordance paradigm. For a 
swift estimate of the parameters the right-hand side of Eq. (\ref{HFnorm4}) can be 
evaluated in the limit $m_{T}(\epsilon) \ll 1$. If we now take the logarithm of both 
sides of Eq. (\ref{HFnorm4}) we obtain
\begin{equation}
[3 - m_{T}(\epsilon)] \, \alpha\, N_{\ast} + \Delta N \, m_{T}(\epsilon) = 15.62 + 15.31\, m_{T}(\epsilon)
\label{HFnorm5}
\end{equation}
In the limit $m_{T}(\epsilon) \ll 1$ Eq. (\ref{HFnorm5}) roughly implies $3\, \alpha \, N_{\ast} \simeq 15.62$.
This mans that $\alpha$ and $N_{\ast}$ are inversely proportional: a longer refractive phase 
imposes a smaller $\alpha$ and vice-versa. Having determined the normalization according 
to Eq. (\ref{HFnorm5}) we can directly write spectral energy density as 
\begin{eqnarray}
h_{0}^2 \Omega_{gw}(\nu) &=& {\mathcal B}(\nu_{L}) \, \biggl(\frac{\nu}{\nu_{\ast}}\biggr)^{m_{T}(\epsilon)}, \qquad\qquad \nu_{\ast}< \nu < \nu_{max},
\nonumber\\
h_{0}^2 \Omega_{gw}(\nu) &=& {\mathcal B}(\nu_{L}) \, \biggl(\frac{\nu}{\nu_{\ast}}\biggr)^{n_{T}(\alpha,\epsilon)}, \qquad\qquad \nu_{eq}< \nu < \nu_{\ast},
\nonumber\\
h_{0}^2 \Omega_{gw}(\nu) &=& {\mathcal B}(\nu_{L}) \, \biggl(\frac{\nu}{\nu_{\ast}}\biggr)^{n_{T}(\alpha,\epsilon)}\,\,\biggl(\frac{\nu_{eq}}{\nu}\biggr)^{2} , \qquad\qquad \nu < \nu_{eq},
\label{HFnorm5a}
\end{eqnarray}
where we stress that ${\mathcal B}(\nu_{L}) = 10^{-8.61}$ has been fixed by normalising 
the spectral energy density to the largest value compatible with the LVK bound. Since 
we already saw  from Fig. \ref{FIG2} that the values of $\nu_{\ast}$ can be either larger 
or smaller than $\nu_{P}$ it is natural to parametrize $\nu_{\ast}$ in terms of $\nu_{P}$ 
by setting $\nu_{\ast} = f_{0} \, \nu_{P}$  where $f_{0}$ can be either smaller or larger 
than $1$. 
\begin{figure}[!ht]
\centering
\includegraphics[height=5.6cm]{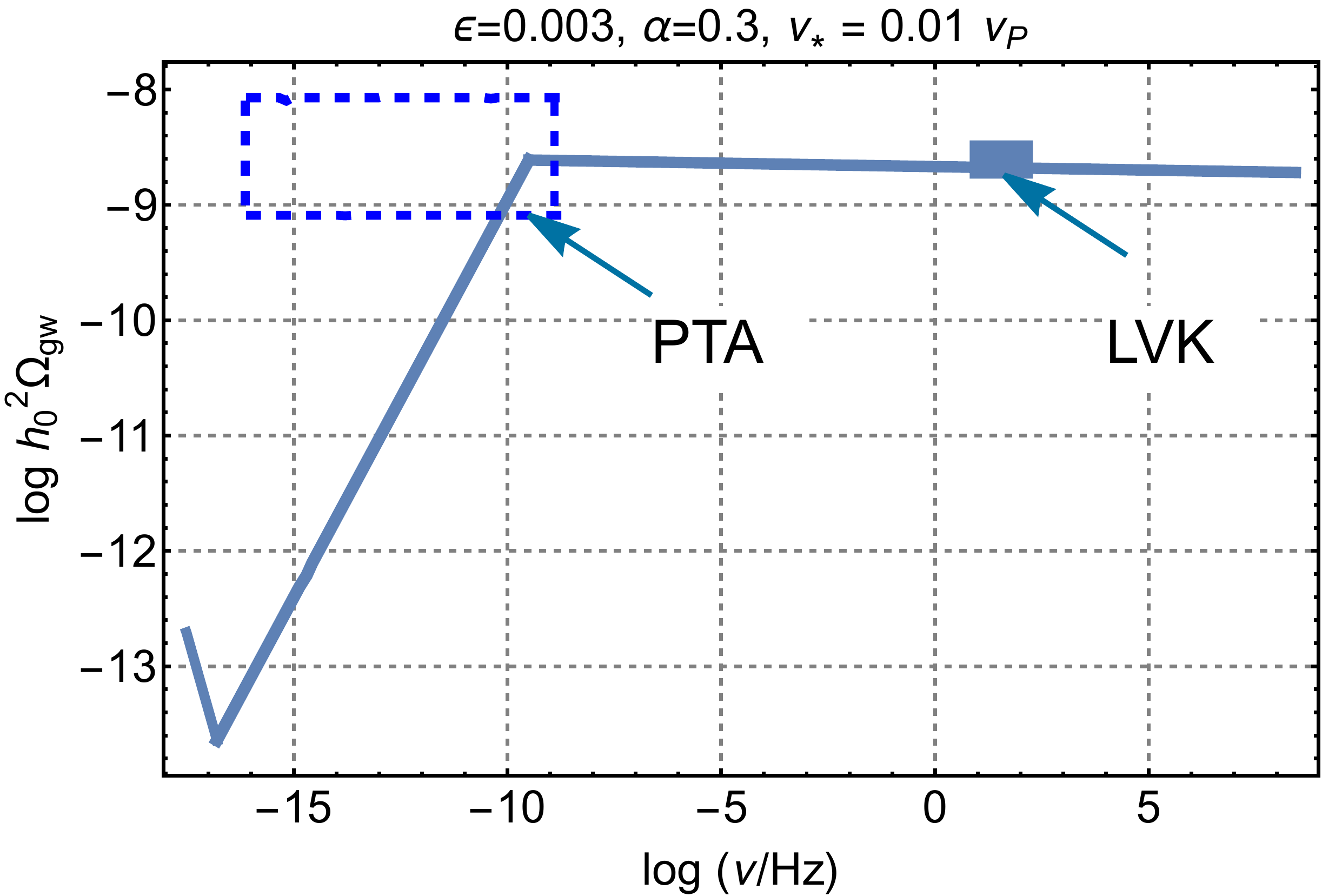}
\includegraphics[height=5.6cm]{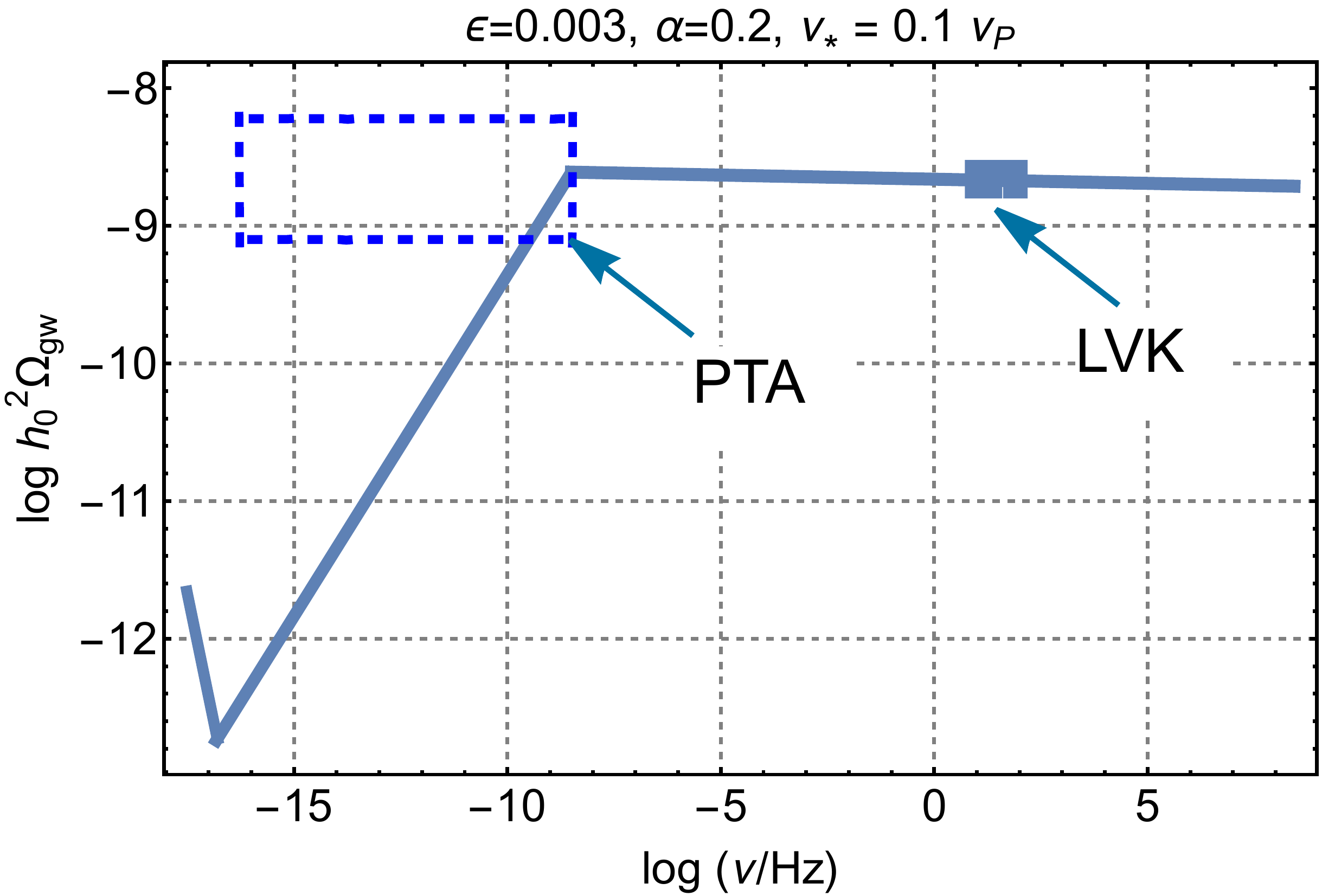}
\caption[a]{The analytic result of Eq. (\ref{HFnorm5}) is illustrated for different values of the parameters. The dashed region defines the purported PTA signal while the LVK limit is automatically taken into account 
by enforcing the high-frequency normalization. We note that these two plots refer to a pair of different 
values of $\nu_{\ast}$ that have been purposely selected around the reference frequency of the PTA which is of the order 
of $31.68$ nHz (see Tab. \ref{TABLE2} and Eq. (\ref{NOTT1})). Note that the region with the dashed perimeter corresponds 
to $q_{0} = \overline{q}_{0} = 2.467$ in Eqs. (\ref{PTAb1}) and (\ref{RESTR2}).}
\label{FIG3}      
\end{figure}
The discussion developed so far assumes a conventional inflationary 
stage where $H_{1}/M_{P}$ is determined from the scalar power spectrum 
and, in this case, Eq. (\ref{HFnorm1}) holds. The same strategy adopted 
above can however be applied also in the complementary 
situation where the universe bounces across a typical curvature scale 
of the order of $H_{1}$. In this case the spectral energy density for $\nu > \nu_{\ast}$ 
is given by: 
\begin{equation}
h_{0}^2 \Omega_{gw}(\nu) = \frac{4 h_{0}^2 \Omega_{R0}}{3 \pi} \biggl(\frac{H_{1}}{M_{P}}\biggr)^2 
\qquad\qquad \nu> \nu_{\ast}.
\label{HFnorm6}
\end{equation}
The main difference is that in the case of Eq. (\ref{HFnorm6}) the scale $H_{1}$ cannot be
estimated as in Eq. (\ref{HFnorm1}). On the contrary it is exactly the LVK bound 
that sets the scale of $H_{1}$:
\begin{equation}
\frac{H_{1}}{M_{P}} = 6.65\times 10^{-3} \biggl(\frac{h_{0}^2 \, \Omega_{R0}}{4.15\times 10^{-5}}\biggr)^{-1/2} \, \biggl(\frac{h_{0}^2 \Omega_{gw}(\nu_{L})}{2.45\times 10^{-9}}\biggr)^{1/2}.
\label{HFnorm7}
\end{equation}
As already mentioned the results of Eqs. (\ref{HFnorm5}) and (\ref{HFnorm6})--(\ref{HFnorm7}) 
do not include the late-time damping that has the effect of shifting a bit the above estimates 
as we shall see in the following section. The logic of this section has been 
to deduce analytically the high-frequency plateau and to normalise its value to the 
LVK bound.

\renewcommand{\theequation}{5.\arabic{equation}}
\setcounter{equation}{0}
\section{Phenomenological considerations}
\label{sec5}
The previous section illustrates how the PTA data 
and the constraints provided by the LVK collaboration set the normalization of the spectral energy density
without any reference to the low-frequency data of the aHz region.
The general expression of the high-frequency 
amplitude consists of two contributions that are formally distinguished 
in Eq. (\ref{OOO}). In the previous section, for the sake of simplicity, 
only the leading contribution to the late-time suppression has been included
while some of the phenomena that may further suppress the high-frequency plateau have been 
neglected. Even though the considerations presented hereunder are relevant,
from a quantitative viewpoint, for an improved determination of the theoretical template, 
the general logic pursued so far remains unaltered. 

 If the late-time contribution (i.e. $\overline{\Omega}_{late}$ in Eq. (\ref{OOO})) is reduced, the early-time contribution 
may become comparatively larger whenever the high-frequency normalization 
is imposed (see also Eqs. (\ref{HFnorm3a})--(\ref{HFnorm3b})). In this respect it is useful to answer three separate
questions. The first one concerns the overall magnitude of  the suppression 
in the high-frequency domain where the normalization is set.
The second issue calls for a distinction between the damping effects 
that depend on the frequency and the ones that are instead frequency-independent. 
It is finally useful to gauge the relative error on the spectral energy density when these 
effects are simply neglected, as tentatively assumed in the previous section.

\subsection{Impact of the neutrino free streaming} 
The effect of neutrino free streaming on the spectral energy density of the relic gravitons has been first 
pointed out by in Ref. \cite{STRESSNU1} (see also \cite{STRESSNU2,STRESSNU3,STRESSNU4,STRESSNU5}) 
where the various authors first argued and then confirmed that the correction to the spectral energy density 
is mildly dependent on the frequency and it is overall of the order of the $10$ \%. 
Since the effect of neutrino free-streaming is fully operational for $\nu <\nu_{bbn}$ it is comparatively less relevant for the high-frequency normalization. The logic is that, after neutrino decoupling, the neutrinos free stream and the effective energy-momentum tensor acquires, to first-order in the amplitude of the plasma fluctuations, an anisotropic stress. In the present case
the spectral energy density at intermediate frequencies is not quasi-flat (as in the standard case) 
but it increases as a function of the frequency but still the magnitude of the effect depends on
 $R_{\nu}$, i.e. the neutrino fraction in the radiation plasma:
\begin{equation}
R_{\nu} = \frac{\rho_{\nu}}{\rho_{\gamma} + \rho_{\nu}} = \frac{3 \times (7/8)\times (4/11)^{4/3}}{ 1 + 3 \times (7/8)\times (4/11)^{4/3}} = 0.4052.
\label{ANIS2d}
\end{equation}
In Eq. (\ref{ANIS2d}) $3$ counts the degrees of freedom associated with the 
massless neutrino families, $(7/8)$ arises because neutrinos follow 
the Fermi-Dirac statistics; the factor  $(4/11)^{4/3}$ stems 
from the relative reduction of the neutrino (kinetic) temperature (in comparison 
with the photon temperature) after weak interactions fall out of thermal 
equilibrium. Assuming that the only collisionless 
species in the thermal history of the Universe are the neutrinos and recalling Eq. (\ref{ANIS2d}), the amount 
of suppression can be parametrized by the function
\begin{equation}
{\mathcal M}(R_{\nu}) = 1 - 0.539 R_{\nu} + 0.134 R_{\nu}^2.
\label{ANIS3}
\end{equation}
This suppression is effective for relatively small frequencies which are larger than $\nu_{eq}$ and smaller than $\nu_{bbn}$.
In what follows we shall stick to the case of the $\Lambda$CDM paradigm \cite{RT1,RT2,RT3} but more 
complicated examples do not change the nature of the effect.

\subsection{Impact of the decoupling of relativistic species}
The late-time suppression of the spectral energy density of Eqs. (\ref{OOO}) and (\ref{HFnorm3a}) has been 
estimated by simply positing that the late Universe is dominated by radiation down to the equality epoch.
If the evolution of the relativistic species is neglected we have 
that $a^4\, H^2$ is roughly constant during radiation. In practice 
to get an approximate expression of $\overline{\Omega}_{late}$ we need 
to evaluate 
\begin{equation}
\frac{a_{re}^4 H_{re}^4}{a_{0}^4 H_{0}^2} = \biggl(\frac{a_{re}^4 H_{re}^4}{a_{eq}^4 H_{eq}^2}\biggr)
 \biggl(\frac{a_{eq}^4 H_{eq}^4}{a_{0}^4 H_{0}^2}\biggr).
 \label{EV1}
 \end{equation}
In the context of the concordance paradigm, the first bracket at the right-hand side of Eq. (\ref{EV1}) 
is evaluated during the radiation stage and it does not introduce any suppression 
as long as  $a^4\, H^2$ is strictly constant. This conclusion is, however,  
slightly inaccurate because of the evolution of the
relativistic species. In local thermal equilibrium 
the total entropy and energy densities of a relativistic 
plasma can be written, respectively, as $s_{t} = 2 \pi^2 g_{s}(T) T^3/(45)$ and 
and as  $\rho_{t} = \pi^2 g_{\rho}(T) T^4/(30)$ where $T$ is the common temperature 
of all the species of the plasma. At the plasma cools down the effective number of spin degrees of freedom 
appearing in the total entropy density and in the total energy density (i.e. $g_{\rho}(T)$ and 
$g_{s}(T)$) eventually decrease in a computable manner depending on the 
microscopic description of the plasma. If all the species of the plasma are in local thermal equilibrium at the 
same temperature we have that $g_{s}$ and $g_{\rho}$ coincide 
and this is what happens in the standard model for temperatures larger 
than the top quark mass where $g_{s}= g_{\rho} =106.75$.
The evolution of the relativistic species suggests
that, during the radiation stage, 
\begin{equation}
\biggl(\frac{a_{r}^4 H_{r}^2}{a^{4} H^2}\biggr) = \biggl(\frac{g_{\rho}(T_{r})}{g_{\rho}(T)}\biggr) \biggl(\frac{g_{s}(T)}{g_{s}(T_{r})}\biggr)^{4/3}.
\label{EFF1a}
\end{equation}
In principle if a given mode $k$ reenters the Hubble radius at a temperature $T_{k}$ the spectral energy density 
of the relic gravitons is (kinematically) suppressed by a factor which can be estimated as $[g_{\rho}(T_{k})/g_{\rho0}] [g_{s}(T_{k})/g_{s0}]^{-4/3}$\cite{extrarel1,extrarel2,extrarel3}; 
at the present time  $g_{\rho0}= 3.36$ and $g_{s0}= 3.90$. In general terms the 
effect parametrized by  Eq. (\ref{EFF1a}) will cause a frequency-dependent suppression, i.e. a further modulation of the spectral energy density $\Omega_{gw}(k,\tau_{0})$.  The maximal suppression one can expect 
follows by inserting into Eq. (\ref{EFF1a}) the largest $g_{s}$ and $g_{\rho}$. 
So, in the case of the minimal standard model this would imply that the suppression (on $\Omega_{gw}(k,\tau_{0})$) will be of the order of ${\mathcal O}(0.38)$. In popular supersymmetric extension of the minimal standard model $g_{\rho}$ and $g_{s}$  can be as large as ${\mathcal O}(230)$ reducing  the previous estimate to ${\mathcal O}(0.29)$. These considerations demonstrate, a posteriori,  the overall correctness of the assumptions behind Eq. (\ref{EFF1a}) and its descendants.

\subsection{Impact of the dominance of dark energy}
The largest wavelengths of the 
spectrum crossed Hubble radius about $65$ $e$-folds before the end of inflation and reentered  when dark energy was already dominant. In the vanilla $\Lambda$CDM scenario the role of the dark energy is played by the cosmological constant and since he evolution of $ \bigl| a\, H \bigr|$ is monotonically decreasing  for $a_{eq} < a < a_{\Lambda}$, all the wavelengths that 
left the  Hubble radius during inflation will remain within the Hubble radius after 
they reenter either during radiation or during matter dominance. If we also consider the evolution for $a > a_{\Lambda}$,  the behaviour of $\bigl|a\, H\bigr|$ is, overall, non-monotonic: a bunch wavelengths  that reentered after equality will again exit after dark energy dominates at the 
redshift 
\begin{equation}
1 + z_{\Lambda} = \biggl(\frac{a_{0}}{a_{\Lambda}}\biggr) = 
 \biggl(\frac{\Omega_{\Lambda}}{\Omega_{\mathrm{M}0}}\biggr)^{1/3}.
\label{LAM1}
\end{equation}
We can now consider  the mode $k_{\Lambda} = a_{\Lambda} H_{\Lambda}$, i.e. the mode 
approximately reentering the Hubble radius at $\tau_{\Lambda}$. Since $H$ is approximately constant for $a > a_{\Lambda}$ we have that $H_{\Lambda} \equiv H_{0}$ where $H_{0}$ is the present value of the Hubble rate.
A typical wavenumber that is today of Hubble size 
(i.e. $k_{0} = a_{0} H_{0}$)  is larger  than $k_{\Lambda}$ even if, according to a superficial 
intuition, it should be smaller\footnote{The numerical difference between $k_{0}$ and $k_{\Lambda}$ is rather insignificant since $k_{\Lambda} = (\Omega_{M0}/\Omega_{\Lambda})^{1/3} k_{0}$ and $(\Omega_{M0}/\Omega_{\Lambda})^{1/3} \simeq {\mathcal O}(0.7)$.  For this reason $k_{0}$, $k_{\Lambda}$ and $k_{p}$ are all coinciding within one order of magnitude. The wavenumbers falling in the interval $k_{0} < k < k_{eq}$ correspond to wavelengths reentering the Hubble radius during the matter-dominated epoch and remaining inside the horizon later on.} and this is because of the non-monotonic evolution of $a H$. This range of wavelengths is
currently inside the Hubble radius and their power spectrum is given by:
\begin{equation}
P_{T}(k, \, \tau_{0}) =P_{T}(k, \tau_{re})  \biggl(\frac{a_{re}}{a_{\Lambda}}\biggr)^2_{\mathrm{mat}} \, 
\biggl(\frac{a_{\Lambda}}{a_{0}}\biggr)^2_{\Lambda} \equiv P_{T}(k, \tau_{re}) 
\biggl(\frac{k_{\Lambda}}{k}\biggr)^{4} \, \biggl(\frac{\Omega_{M0}}{\Omega_{\Lambda}}\biggr)^{2/3},
 \qquad k_{0} < k < k_{eq}.
\label{LAM2}
\end{equation}
The spectral energy density when the relevant wavelengths are inside the Hubble radius can be obtained from Eq. (\ref{LAM2}) and the result is:
\begin{eqnarray}
\Omega_{gw}(k,\tau_{0}) &=& \frac{P_{T}(k,\tau_{re})}{12} \biggl(\frac{k}{k_{0}}\biggr)^{-2} \biggl(\frac{k_{\Lambda}}{k_{0}}\biggr)^4 \biggl(\frac{\Omega_{M0}}{\Omega_{\Lambda}}\biggr)^{2/3} 
\nonumber\\
&=&  \frac{P_{T}(k,\tau_{re})}{12}\, \biggl(\frac{k}{k_{0}}\biggr)^{-2} \,  \biggl(\frac{\Omega_{M0}}{\Omega_{\Lambda}}\biggr)^{2}, \qquad k_{0} < k < k_{eq},
\label{LAM4}
\end{eqnarray}
where we used that, by definition, $k_{\Lambda}/k_{0} = (\Omega_{M0}/\Omega_{\Lambda})^{1/3}$. 
If the dominance of dark energy is completely neglected,  Eq. (\ref{LAM4}) can be written in the same 
form where, however, the term $(\Omega_{M0}/\Omega_{\Lambda})^2$ is absent. We conclude that, in this 
branch of the spectrum, the dominance of dark energy suppresses the spectrum by a factor 
$(\Omega_{M0}/\Omega_{\Lambda})^2 = (0.44)^2 = 0.193$. 

The wavenumbers falling instead in the interval $k_{\Lambda} < k < k_{0}$ correspond to wavelengths reentering the Hubble radius during the matter-dominated 
epoch and exiting again the Hubble radius when dark energy is already dominant. These wavelengths are currently outside the Hubble radius. The
same logic leading  to Eq. (\ref{LAM2}) implies that the power spectrum in the interval
$k_{\Lambda} < k < k_{0}$
is given by $ P_{T}(k,\, \tau_{re})\, (a_{re}/a_{ex})^2$ or, more precisely, 
\begin{equation}
P_{T}(k, \, \tau_{0}) = P_{T}(k, \tau_{re})  \biggl(\frac{a_{re}}{a_{\Lambda}}\biggr)^2_{\mathrm{mat}} \, 
\biggl(\frac{a_{\Lambda}}{a_{ex}}\biggr)^2_{\Lambda} \equiv P_{T}(k, \tau_{re}) 
\biggl(\frac{k_{\Lambda}}{k}\biggr)^{6},
 \qquad k_{\Lambda} < k < k_{0}.
\label{LAM5}
\end{equation}
\begin{figure}[!ht]
\centering
\includegraphics[height=7.5cm]{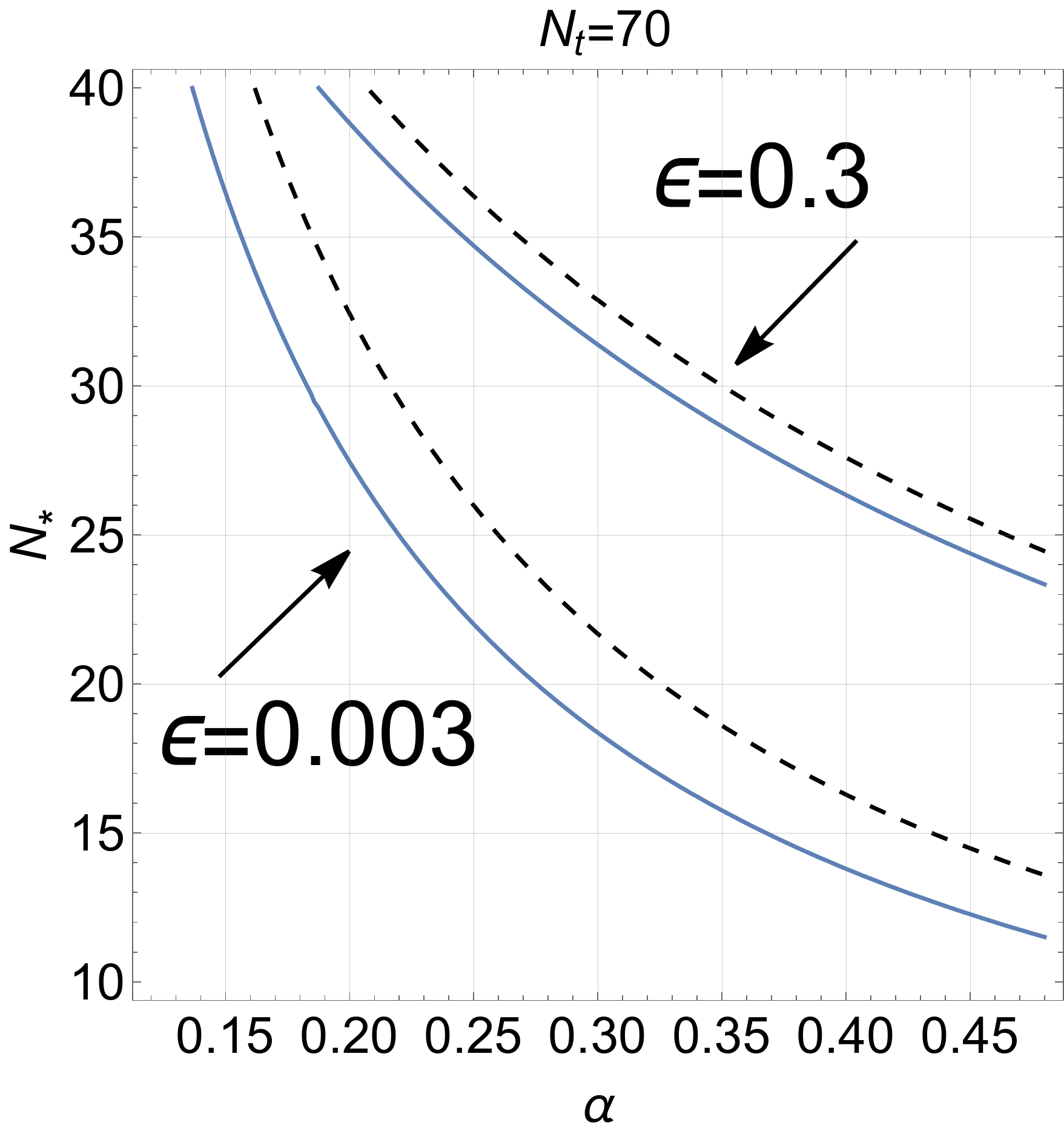}
\caption[a]{In the plane $(\alpha, \, N_{\ast})$ we illustrate the normalization curve of Eq. (\ref{HFnorm4}).
The dashed lines include the damping effects discussed in this section. The full line comes instead 
only from the estimates of the previous section. For the sake of illustration we selected $N_{t}=70$ 
but different values for the total number of $e$-folds do not change the main features 
of the plot.}
\label{FIG4}      
\end{figure}
Equation (\ref{LAM5}) gives the power spectrum for typical wavelengths that are larger 
than the Hubble radius at the present time. The wavelengths belonging to the interval 
$k_{\Lambda} < k < k_{0}$ reenter the Hubble radius but then exit again and while they are 
larger than the Hubble radius their power spectrum remains unchanged. This
is why, in Eq. (\ref{LAM5}), the power spectrum is not suppressed by a further factor 
$(a_{re}/a_{0})^2$: these wavelengths are larger than the Hubble radius at the present time 
and therefore the power spectrum remains constant exactly as in the case of the scales 
exiting the Hubble radius at the onset of inflation. Using the identity $k_{\Lambda}/k_{0} = (\Omega_{M0}/\Omega_{\Lambda})^{1/3}$ the spectrum of Eq. (\ref{LAM5}) can be written as:
\begin{equation}
P_{T}(k, \, \tau_{0}) = P_{T}(k, \tau_{re}) \biggl(\frac{k_{0}}{k}\biggr)^{6} \biggl(\frac{\Omega_{M0}}{\Omega_{\Lambda}}\biggr)^2.
\label{LAM6}
\end{equation}
We can finally compute the spectral energy density in this interval and we obtain 
\begin{eqnarray}
\Omega_{gw}(k,\tau_{0}) = \frac{P_{T}(k,\tau_{re})}{12}\, \biggl(\frac{k}{k_{0}}\biggr)^{-4} \,  \biggl(\frac{\Omega_{M0}}{\Omega_{\Lambda}}\biggr)^{2},  \qquad k_{\Lambda} < k < k_{0}.
\label{LAM7}
\end{eqnarray}
If the presence of dark energy would be neglected, $\Omega_{gw}(k,\tau_{0})$ would 
scale as $(k/k_{0})^{-2}$ and the suppression going as $(\Omega_{M0}/\Omega_{\Lambda})^{2}$ 
would be totally absent. Note finally that when $k = k_{\Lambda}$ Eq. (\ref{LAM6}) implies that 
$\Omega_{gw}(k_{\Lambda},\tau_{0})$ is only suppressed as $(\Omega_{M0}/\Omega_{\Lambda})^{4/3}$.

\subsection{Numerical discussion of the spectral energy density}
After having assessed the main sources of damping at high-frequency we now go back 
to Eqs. (\ref{HFnorm3a})--(\ref{HFnorm3b}) and (\ref{HFnorm4}) with the purpose of improving the 
quantitive evaluation of the late-time suppression. Since 
the neutrino free-streaming is only effective at comparatively low-frequencies, at the right-hand 
side of Eq. (\ref{HFnorm4}) we should add a further contribution, namely 
\begin{equation}
  \ln{\biggl[\frac{g_{\rho}(T)}{g_{\rho}(T_{r})}\biggr]} + \frac{4}{3} \ln{\biggl[\frac{g_{s}(T_{r})}{g_{s}(T)}\biggr]} - 2 \ln{\biggl(\frac{\Omega_{M0}}{\Omega_{\Lambda}}\biggr)}
= {\mathcal O}(3).
\label{EV2}
\end{equation}
Equation (\ref{EV2}) gives in fact the maximal suppression in the audio band 
so that, assuming $m_{T}(\epsilon) \ll 1$ the condition (\ref{HFnorm4}) gets modified as 
$3 \alpha N_{\ast} \simeq 18.62$. For the same reason the determination of Eq. (\ref{HFnorm7}) 
gets is also affected and it is now
\begin{equation}
\frac{H_{1}}{M_{P}} = 1.15\times 10^{-2} \biggl(\frac{h_{0}^2 \, \Omega_{R0}}{4.15\times 10^{-5}}\biggr)^{-1/2} \, \biggl(\frac{h_{0}^2 \Omega_{gw}(\nu_{L})}{2.45\times 10^{-9}}\biggr)^{1/2}.
\label{EV3}
\end{equation}
In Fig. \ref{FIG4} we illustrate the normalization curves deduced in Eq. (\ref{HFnorm4}). In particular, while  the full lines 
refer to the case when the late-time damping is neglected, the dashed 
curves include instead the different sources of damping previously analyzed in this section.
The contribution associated with the neutrino free-streaming affects frequencies that are 
approximately smaller than $\nu_{bbn}$ and, for this reason, they are less relevant for the high-frequency 
normalization. At intermediate frequencies the spectral energy density is instead 
affected and this observation is relevant for the potential signal of the PTA. 
\begin{figure}[!ht]
\centering
\includegraphics[height=8.3cm]{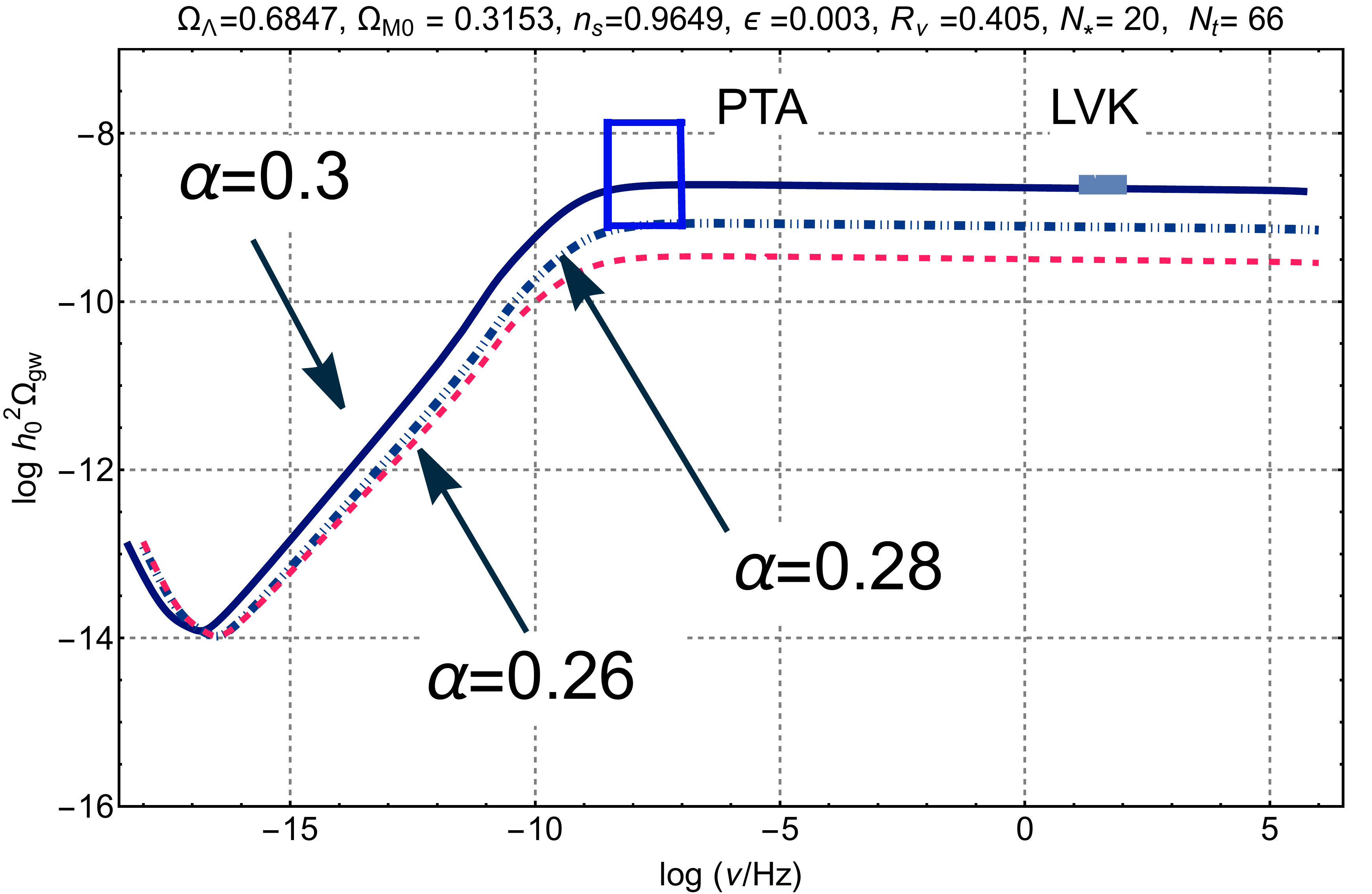}
\caption[a]{We illustrate the spectral energy density for different values 
of $\alpha$. Common logarithm are employed on both axes. The effect of neutrino free-streaming is responsible for the suppression at intermediate frequencies. For the sake of illustration we consider $\alpha = {\mathcal O}(0.28)$ since $\alpha =2/7$ ultimately corresponds to $\beta \to 2/3$ in Eqs. (\ref{NOTT1})--(\ref{NOTT8}), and this is the value presumably suggested by the PTA determinations (see also Tab. \ref{TABLE2}). Note that the PTA region corresponds 
to $q_{0} = \overline{q}_{0} = 2.467$ in Eqs. (\ref{PTAb1}) and (\ref{RESTR2}).}
\label{FIG5}      
\end{figure}
This aspect is illustrated in Fig. \ref{FIG5} where the spectral energy density has been reported for an illustrative choice 
of the parameters that is now specifically discussed. The pivotal parameters that determine the spectrum are primarily $\alpha$, $N_{\ast}$ and $N_{t}$. When $N_{*}$ and $N_{t}$ are of the same order the transition to normalcy occurs at the end of inflation
but in this case it is impossible to get a large signal in the nHz range without jeopardizing the big-bang nucleosynthesis constraint of Eqs. (\ref{CC2})--(\ref{CC3}).  If we ought to address the PTA measurements  we must require  $N_{*} <N_{t}$ since, in this case, the transition to normalcy takes place before the onset of the radiation-dominated epoch (i.e. when the background is still inflating deep inside the quasi-de Sitter stage of expansion). In Fig. \ref{FIG5}  we choose $N_{t} =65$ and set $\epsilon=0.003$. Even if this value 
has been deduced from the upper limits on $r_{T}$ \cite{RT1,RT2,RT3} and from the consistency we could 
easily consider values $\epsilon \ll 0.003$ and $r_{T} \ll 0.06$ since they are immaterial for the overall 
normalization but only for the slope $m_{T}(\epsilon)$ of the high-frequency plateau. The values of $\alpha$ appearing 
in Fig. \ref{FIG5} are around $\alpha \simeq 2/7$. In this case,  the spectral index at intermediate frequencies is given by $n_{T} \simeq 2/3 $, up to slow-roll corrections ${\mathcal O}(\epsilon)$ which are negligible since $\epsilon < 10^{-3}$. This 
 value of $\alpha$ actually corresponds to $\beta=-2/3$ (see, in this respect, Tab. \ref{TABLE2} and discussions thereafter). As discussed the PTA results are often reported in terms of a chirp amplitude scaling as $\nu^{-2/3}$ for a typical reference frequency ${\mathcal O}(\mathrm{yr}^{-1})$. The value $\alpha =2/7 \simeq 0.28$ corresponds to $\beta= -2/3$ and  $n_{T} \simeq 2/3$. More precisely we have that $\alpha = (2 + 4 \epsilon)/7$ which can be approximated as $\alpha = 2/7 + {\mathcal O}(\epsilon)$ 
for $\epsilon < 10^{-3}$. In Fig. \ref{FIG5} the box illustrates the PTA measurements. In 
Fig. \ref{FIG5} the spectral energy density in critical units has been normalized by using the limits from the 
wide-band detectors given in Tab. \ref{TABLE1} and by also imposing, as a particular choice, that $\alpha = {\mathcal O}(0.28)$
as suggested by the PTA measurements of Tab. \ref{TABLE2}. It is interesting that these 
two independent choices lead to a large signal in the nHz range when the variation of the refractive index occurs sufficiently early during the inflationary stage and anyway not beyond the first $20$ $e$-folds. It is finally important 
to remark that, in the present context, then constraints on the integral of the spectral energy density of the relic gravitons (see Eqs. (\ref{CC2})--(\ref{CC3})) are automatically satisfied after imposing the normalization deduced from the direct 
limits set by the operating interferometers (see Tab. \ref{TABLE1} and discussion thereafter).
 
\renewcommand{\theequation}{6.\arabic{equation}}
\setcounter{equation}{0}
\section{Concluding remarks}
\label{sec6}
Within the concordance paradigm the absolute normalization of the spectral energy 
density of the relic gravitons in critical units is assigned in the aHz region and it depends on  $r_{T}$ (i.e. the tensor to scalar ratio) that should not exceed ${\mathcal O}(0.06)$, at least according to the limits set by the temperature and polarization anisotropies of the microwave background. Since the $\Lambda$CDM is a compromise between the available data and the number of ascertainable parameters, the most stringent limits on $r_{T}$  hold when the consistency relations between the scalar and tensor power spectra are enforced; in practice this only happens in the case of single-field inflationary models. If the spectral energy density of the cosmic gravitons is predominantly distributed for frequencies much larger than the aHz the logic leading to the low-frequency normalization is less compelling. In the nHz region the PTA recently reported a potential excess even if a bona fide signal coming from the relic gravitons should be correlated across the baselines and so far no indications along this direction have been obtained. Motivated by the improved limits in the audio band and by the current data from the pulsar timing arrays in the nHz domain, we analyzed the conditions for a quasi-flat spectrum of relic gravitons at intermediate and high-frequencies by introducing an improved physical strategy for the absolute normalization of the cosmic background of relic gravitons. 
 
After proposing a general four-dimensional action for the discussion of relic gravitons 
in spatially flat backgrounds, we concentrated on the classes of scenarios
where a large signal can be expected between the nHz and kHz ranges. While the most promising possibilities involve a dynamical refractive index and the bouncing dynamics, between these two cases the former is slightly more conservative than the latter insofar as it is compatible with the presence adiabatic and Gaussian initial data for the temperature and polarisation anisotropies of the microwave background. In both situations the 
spectral energy density increases over intermediate frequencies and then flattens out.  
Since the data from the wide-band detectors set the normalization 
of the spectral energy density at high-frequencies, a scheme based on the WKB method is preferable for a general estimate. Within this approach the early contributions can be  easily distinguished from the late-time effects that are evaluated, depending on the convenience, in different approximations.  The results obtained here also suggest an effective mechanism for the origin of a flat spectrum of relic gravitons with typical amplitudes that are even six or seven orders of magnitude larger than in the case of conventional inflationary models. 

The results obtained here also suggest that the signal of the PTA 
can be explained by a background of relic gravitons of inflationary origin without 
conflicting with the bounds coming from big-bang nucleosynthesis which are automatically satisfied as long as the data from wide-band interferometers set the normalization of the spectral energy density in critical units between few Hz and $0.1$ kHz. In the present framework the low-frequency constraints can also be imposed a posteriori as a limit on the intermediate spectral slope but they are overall less crucial and they should be applied, strictly speaking, only when the consistency conditions between scalar and tensor modes are enforced.

\section*{Acknowledgements}

The author is indebted to T. Basaglia, A. Gentil-Beccot, S. Rohr and J. Vigen of the CERN Scientific Information Service for their kind assistance.


\newpage

\begin{thebibliography}{99}

\bibitem{gr1} L.~P.~Grishchuk,   Sov.\ Phys.\ JETP {\bf 40}, 409 (1975)   [Zh.\ Eksp.\ Teor.\ Fiz.\  {\bf 67}, 825 (1974)].

\bibitem{gr2}  L.~P.~Grishchuk,   Annals N.\ Y.\ Acad.\ Sci.\  {\bf 302}, 439 (1977).

\bibitem{par1} L.~H.~Ford and L.~Parker,  Phys.\ Rev.\ D {\bf 16}, 1601 (1977).

\bibitem{star1} A.~A.~Starobinsky, JETP Lett.\  {\bf 30}, 682 (1979) [Pisma Zh.\ Eksp.\ Teor.\ Fiz.\  {\bf 30}, 719 (1979)]. 

\bibitem{RT1}  Y.~Akrami {\it et al.} [Planck Collaboration], Astron. Astrophys. {\bf 641}, A10 (2020).

\bibitem{RT2}  N.~Aghanim {\it et al.} [Planck Collaboration], Astron. Astrophys. {\bf 641}, A6 (2020).

\bibitem{RT3} P.~A.~R.~Ade {\it et al.} [BICEP and Keck], Phys. Rev. Lett. {\bf 127}, 151301 (2021).

\bibitem{ir1} V. A. Rubakov, M. V. Sazhin and A. V. Veryaskin, Phys. Lett. {\bf 115B}, 189 (1982).

\bibitem{ir2} R.~Fabbri and M.~D.~Pollock, Phys.\ Lett.\  {\bf 125B}, 445 (1983).

\bibitem{ir3} L. F.  Abbott and M. B. Wise, Nucl. Phys. {\bf 224}, 541 (1984).

\bibitem{wein} S.~Weinberg, {\it Cosmology}, (Oxford Univ. Press, Oxford UK,  2008).

\bibitem{norm} M.~Giovannini, Phys. Lett. B \textbf{668}, 44 (2008).

\bibitem{PUL1} V.~M.~Kaspi, J.~H.~Taylor, and M.~F.~Ryba,   Astrophys.\ J.\ {\bf 428}, 713 (1994).

\bibitem{PUL2} F.~A.~Jenet {\it et al.},  Astrophys.\ J.\  {\bf 653}, 1571 (2006).

\bibitem{PUL3} W.~Zhao,  Phys.\ Rev.\ D {\bf 83}, 104021 (2011).

\bibitem{PUL4} P.~B.~Demorest {\it et al.},  Astrophys.\ J.\  {\bf 762}, 94 (2013).

\bibitem{PUL5}  W.~Zhao, Y.~Zhang, X.~P.~You and Z.~H.~Zhu,  Phys.\ Rev.\ D {\bf 87},  124012 (2013).

\bibitem{NANO1} Z. Arzoumanian {\it et al.}, Astrophys. J. Lett. {\bf 905}, L34 (2020).

\bibitem{PPTA} B. Goncharov {\it et al.} Astrophys. J. Lett. {\bf 917}, L19 (2021).

\bibitem{EPTA} S.~Chen, {\it et al.} Mon. Not. Roy. Astron. Soc. {\bf 508},  4970 (2021). 

\bibitem{IPTA} J.~Antoniadis {\it et al.}, [arXiv:2201.03980 [astro-ph.HE]].

\bibitem{REFR1}  M.~Giovannini, Class.\ Quant.\ Grav.\  {\bf 33}, 125002 (2016).

\bibitem{REFR2} M.~Giovannini, Phys. Rev. D {\bf 98}, 103509 (2018).

\bibitem{STone}   B.~Abbott {\it et al.} [LIGO Collaboration], Phys.\ Rev.\ D {\bf 69}, 122004 (2004).
  
\bibitem{STonea} B.~Abbott {\it et al.} [LIGO Collaboration], Phys.\ Rev.\ Lett.\  {\bf 95}, 221101 (2005).  
  
 \bibitem{STtwo}  J.~Abadie {\it et al.} [LIGO/Virgo Collaboration], Phys.\ Rev.\ D {\bf 85}, 122001 (2012).

\bibitem{STthree}  J.~Aasi {\it et al.} [LIGO/Virgo Collaboration], Phys.\ Rev.\ Lett.\  {\bf 113}, 231101 (2014).

\bibitem{STthreea} J.~Aasi {\it et al.} [LIGO/Virgo Collaboration],  Phys.\ Rev.\ D {\bf 91},  022003 (2015).
  
\bibitem{STfour}  B.~P.~Abbott {\it et al.} [LIGO/Virgo Collaboration],  Phys.\ Rev.\ Lett.\  {\bf 118}, 121101 (2017)  Erratum: [Phys.\ Rev.\ Lett.\  {\bf 119}, 029901 (2017)].

\bibitem{STfive}  B.~P.~Abbott {\it et al.} [LIGO/Virgo Collaboration],Phys.\ Rev.\ D {\bf 100}, 061101(R) (2019).

\bibitem{STsix} R.~Abbott \textit{et al.} [LIGO, Virgo and KAGRA collaborations], Phys. Rev. D {\bf 104},  022004 (2021).

\bibitem{ONEW} S. Weinberg, Phys. Rev. D {\bf 77}, 123541 (2008).

\bibitem{NON1} H.~Motohashi and A.~A.~Starobinsky, JCAP {\bf 11}, 025 (2019).

\bibitem{NON2} M.~Guerrero, D.~Rubiera-Garcia and D.~Saez-Chillon Gomez, Phys. Rev. D {\bf 102}, 123528 (2020).

\bibitem{NON3} A.~Mohammadi, T.~Golanbari, S.~Nasri and K.~Saaidi, Phys. Rev. D {\bf 101},  123537 (2020).

\bibitem{NON4} M.~Gasperini and M.~Giovannini, Phys. Lett. B {\bf 287}, 56 (1992).

\bibitem{NON5} I.~Antoniadis, J.~Rizos and K.~Tamvakis, Nucl. Phys. B {\bf 415}, 497 (1994).

\bibitem{NON6} Z. Guo and D. Schwarz, Phys. Rev. D {\bf 80}, 063523 (2009).

\bibitem{NON7} S.~M.~Carroll and E.~A.~Lim, Phys. Rev. D \textbf{70}, 123525 (2004).

\bibitem{NON8} W.~Donnelly and T.~Jacobson, Phys. Rev. D {\bf 82}, 081501 (2010).

\bibitem{NON9} J.~D.~Barrow, Phys. Rev. D \textbf{85}, 047503 (2012).

\bibitem{CC1a} P.~Szekeres,  Annals Phys.\  {\bf 64}, 599 (1971).

\bibitem{CC1b} P.~C.~Peters,  Phys.\ Rev.\ D {\bf 9}, 2207 (1974).

\bibitem{PPNN} M.~Giovannini, Eur. Phys. J. C {\bf 82},  117 (2022). 

\bibitem{DEC1a} M.~Giovannini, Phys. Rev. D \textbf{55}, 595 (1997).

\bibitem{DEC2a} M.~Giovannini, Class. Quant. Grav. \textbf{22}, 2201 (2005).

\bibitem{TWO}  S.-Y. Pi and R. Jackiw, Phys. Rev. D {\bf 68}, 104012 (2003).

\bibitem{pol} M.~Giovannini, Phys. Rev. D \textbf{99}, 083501 (2019).

\bibitem{SN1} N. Christensen, Phys. Rev. D {\bf 55}, 448 (1997).

\bibitem{SN2} B. Allen and J. Romano, Phys. Rev. D {\bf 59}, 102001 (1999).

\bibitem{SN3} D.~Babusci and M.~Giovannini,  Phys.\ Rev.\ D {\bf 60}, 083511 (1999).

\bibitem{SN4} D.~Babusci and M.~Giovannini,  Class.\ Quant.\ Grav.\  {\bf 17}, 2621 (2000).

 \bibitem{abr1} M. Abramowitz and I.A. Stegun, {\it Handbook of Mathematical Functions} (Dover, New York, 1972).

\bibitem{abr2} A. Erdelyi, W. Magnus, F. Obehettinger, and F. Tricomi, {\it Higher Trascendental Functions} (McGraw-Hill, New York, 1953).

\bibitem{bbn1} V.~F.~Schwartzmann, JETP Lett.\ {\bf 9}, 184 (1969).
  
\bibitem{bbn2} M.~Giovannini, H.~Kurki-Suonio and E.~Sihvola, Phys.\ Rev.\  D {\bf 66}, 043504 (2002).

\bibitem{bbn3} R.~H.~Cyburt, B.~D.~Fields, K.~A.~Olive, and E.~Skillman, Astropart.\ Phys.\ {\bf 23}, 313 (2005).

\bibitem{STRESSNU1}   S.~Weinberg,  Phys.\ Rev.\  D {\bf 69}, 023503 (2004).

\bibitem{STRESSNU2} D.~A.~Dicus and W.~W.~Repko,  Phys.\ Rev.\  D {\bf 72}, 088302 (2005).
 
\bibitem{STRESSNU3} H.~X.~Miao and Y.~Zhang, Phys.\ Rev.\ D {\bf 75}, 104009 (2007).
 
\bibitem{STRESSNU4}  K.~W.~Ng, Phys.\ Rev.\ D {\bf 86}, 103510 (2012).
 
 \bibitem{STRESSNU5}  B.~A.~Stefanek and W.~W.~Repko, Phys.\ Rev.\ D {\bf 88},  083536 (2013).
 
 \bibitem{extrarel1}  Y.~Watanabe and E.~Komatsu,  Phys.\ Rev.\  D {\bf 73}, 123515 (2006).

\bibitem{extrarel2} W.~Zhao and Y.~Zhang,  Phys.\ Rev.\  D {\bf 74}, 043503 (2006).

\bibitem{extrarel3} K.~Saikawa and S.~Shirai,  JCAP {\bf 1805}, 05, 035 (2018).


\end{thebibliography}
\end{document}